\documentclass[a4paper,11pt]{article}
\usepackage{jheppub}

\usepackage[english]{babel}
\usepackage{amsmath,amssymb}

\newcommand{\tmop}[1]{\ensuremath{\operatorname{#1}}}

\preprint{Imperial-TP-2024-CH-7}

\title{Monopoles, Dirac Strings and Generalised Symmetries}

\author{C M Hull}
\affiliation{The Blackett Laboratory, Imperial College London, Prince Consort Road, London, SW7 2AZ, UK}
\emailAdd{c.hull@imperial.ac.uk}

\abstract
{
Dirac's formulation of magnetic monopoles is shown to be equivalent to Maxwell theory
coupled to 2-form gauge fields so that it has a local 1-form symmetry, with
the  2-form gauge fields given in terms of the 2-form current densities associated with the Dirac strings. The field equations of Dirac's theory do not depend on the positions of the Dirac strings provided that they do not intersect the worldlines of any electrically charged particles; this constraint is called the Dirac veto.  It is shown that Dirac's action
is   independent of the positions of the Dirac strings and that this corresponds to a local 1-form symmetry. The electric and magnetic 1-form symmetries have a mixed anomaly, and the Dirac veto is shown to correspond to a restriction to gauge transformations for which the anomaly vanishes. 
The extension to $p$-form gauge fields in $d$-dimensions coupled to charged branes is discussed, together with the possibility of cancelling the anomaly by embedding in a higher-dimensional theory and so avoiding the veto.
 }

\begin{document}
\maketitle

\flushbottom

\section{Introduction and summary}

In his 1948 paper \cite{Dirac:1948um}, Dirac gave a quantum theory of the electromagnetic field in 4
dimensions coupling to both electrically charged particles and magnetic monopoles. The 2-form field strength $F$
 satisfies the field equations
 \begin{equation}
  d^{\dag} F = j
\end{equation}
\begin{equation}
  d F = \ast \tilde{j} \label{max2}
\end{equation}
where the sources are
 an electric 1-form current $j$ and a magnetic
1-form current $\tilde{j}$, both of which are conserved, ${d }^{\dag}j={d }^{\dag} \tilde{j} =0$.\footnote{In this
introduction the focus will be on 4-dimensional Minkowski space, but much of
the formalism extends to general spacetimes. In 4 dimensions with Lorentzian
signature, the Hodge dual satisfies $\ast^2 = - 1$ when acting on forms of
even degree and $\ast^2 = 1$ on forms of odd degree, while $d^{\dag} = \ast d
\ast$.}
Dirac solved the equation (\ref{max2}) by introducing a 2-form current
$\tilde{J}$ satisfying
\begin{equation}
  \begin{array}{lll}
    {d }^{\dag} \tilde{J} & = & \tilde{j}
  \end{array} \label{Jjt}
\end{equation}
so that
\begin{equation}
  d (F - \ast \tilde{J}) = 0
\end{equation}
and there is a $1$-form potential $A$ satisfying
\begin{equation}
  F = \ast \tilde{J} + d A \label{fja}
\end{equation}
The $1$-form $\tilde{j}$ is the magnetic monopole current. A Dirac string is
attached to each magnetic monopole and the $2$-form $\tilde{J}$ is the current
density for these strings: if  $\tilde{j}$ is localised on the world-line of
a magnetic monopole, then $\tilde{J}$ is localised on the world-sheet of the
corresponding Dirac string.

  Dirac's action is given by the sum of the kinetic terms for the electric
and magnetically charged particles plus
\begin{equation}
  S [A] = \int \frac{1}{2} F \wedge \ast F - A \wedge \ast j \label{act}
\end{equation}
with $F = \ast \tilde{J} + d A$. This gives the correct field equations,
provided that the condition that has become known as the
{\emph{Dirac veto}} holds. This requires that the
positions of the Dirac strings be restricted so that there is no intersection
between the world-lines of electrically charged particles and the world-sheets
of the Dirac strings. In particular, the field equations do not depend on the
locations of the Dirac strings provided that they comply with the Dirac veto,
and so do not depend on the choice of $\tilde{J}$ satisfying (\ref{Jjt}).
If there are no magnetic sources then $\tilde{j}=0$ and $\tilde{J}=0$ and the theory reduces to the usual Maxwell action.

Dirac's action is not single-valued. A continuous deformation of the
positions of the Dirac strings (while obeying the veto) can change the action by any integral multiple
of $4 \pi q p$ where $q$ is the electric charge of any particle and $p$ is the
magnetic charge of any monopole \cite{Dirac:1948um}. Then $e^{i S / \hbar}$ will be single
valued provided the electric and magnetic charges all satisfy the Dirac
quantisation condition and the quantum theory is then well-defined.

As Dirac pointed out \cite{Dirac:1948um}, Dirac strings can instead be introduced  for the
electrically charged particles. If the electric current is $j$, there is then
a $2$-form current $J$ localised on the world-sheets of the electric Dirac
strings satisfying
\begin{equation}
  {d }^{\dag} J = j
\end{equation}
Then
\[ d^{\dag} F = j \]
can be written as
\begin{equation}
  d (\tilde{F} - \ast J) = 0
\end{equation}
(writing $\tilde{F} = \ast F$ for the Hodge dual of $F$) so that there is a dual
formulation in terms of a dual potential $\tilde{A}$ with
\begin{equation}
  \tilde{F} = \ast J + d \tilde{A}
\end{equation}
with action
\begin{equation}
  S [\tilde{A}] = \int \frac{1}{2} \tilde{F} \wedge \ast \tilde{F} - \tilde{A}
  \wedge \ast \tilde{j} \label{duact}
\end{equation}
In this case, Dirac's veto requires that the electric Dirac strings on which
the current $J$ is localised do not intersect the world-lines of the
magnetically charged particles; this will be referred to as the dual Dirac veto.

Returning to the original formulation with magnetic Dirac strings, $\tilde{J}$
is singular on the world-sheets of the Dirac strings and zero
elsewhere.\footnote{The singularity is a delta-function singularity. These can
be regularised by replacing delta functions with smooth bump functions that approach
delta functions in a suitable limit, as in \cite{Dirac:1948um}, so that sources are smooth
currents rather than delta-functions.} For $F = \ast \tilde{J} + d A$ to be a
non-singular 2-form, it must be that $A$ is also singular on the Dirac strings
and non-singular elsewhere. The term $- \int A \wedge \ast j$ in the action is
then well-defined as the Dirac veto ensures that $A$ is non-singular on the
electric particle world-lines where $j$ is non-zero. An alternative action is
given by replacing this term with $\int F \wedge \ast J$. The two terms differ
by a surface term; whether or not this vanishes depends on the boundary
conditions. The action
\begin{equation}
  \hat{S} [A] = \int \frac{1}{2} F \wedge \ast F + F \wedge \ast J
  \label{acth}
\end{equation}
then gives the same field equations as (\ref{act}).

Dirac added  kinetic terms for the electrically and magnetically charged particles
to the  action for the electromagnetic field, so that the currents $j,\tilde j$ arise dynamically. He went on to quantise the system and so provide a  quantum theory for the electromagnetic field coupled to quantum  electrically and magnetically charged particles.

Dirac showed that his field equations do not depend on the position of the
Dirac strings, provided that  they comply with the Dirac veto. 
In
 this paper,
 the dependence of the action on the position of the strings will be investigated. 
 It will be argued that, in fact, the action  is invariant under changing the
positions of the Dirac strings (subject to the Dirac veto)
 and that this invariance can be formulated  in terms of  extra gauge symmetries of the action.
These are 1-form
generalised symmetries and it will be seen that the Dirac veto arises as a
condition for the absence of anomalies in the generalised 1-form symmetries.

The starting point is to  note that the equations
\begin{equation}
  {d }^{\dag} J = j,  \quad \begin{array}{lll}
    {d }^{\dag} \tilde{J} & = & \tilde{j}
  \end{array}
\end{equation}
don't determine the currents $J, \tilde{J}$ uniquely, but only determine them
up to co-closed terms. In particular, they can be transformed by
\begin{equation}
  \delta J = d^{\dag} \rho, \quad \delta \tilde{J} = d^{\dag} \tilde{\rho}
  \label{Jro}
\end{equation}
for some 3-forms $\rho, \tilde{\rho}$. The interpretation of these transformations  is
as follows. Smoothly deforming the 2-surface $\mathcal{D} $ on which a Dirac
string is localised to a 2-surface $\mathcal{D} '$ gives a family of Dirac
string world-sheets $\mathcal{D} (\xi)$ parameterised by $\xi\in [0,1]$ with
$\mathcal{D} (0) =\mathcal{D}$ and $\mathcal{D} (1) =\mathcal{D}'$. This
family of 2-dimensional world-sheets sweeps out a 3-dimensional surface
$\mathcal{E}$. For a magnetic Dirac string, the resulting change in $\tilde{J}
$ is, for an infinitesimal deformation of $\mathcal{D}$, of the form $\delta
\tilde{J} = d^{\dag} \tilde{\rho}$ where $\tilde{\rho}$ is a 3-form current 
localised on $\mathcal{E}$. This will be discussed in more detail in section
5. Similarly, for a deformation of an electric Dirac string the change in
$J$ is $\delta J = d^{\dag} \rho$ where $\rho$ is a 3-form current localised
on $\mathcal{E}$.

In order for $F = \ast \tilde{J} + d A$ and $\tilde{F} = \ast J + d
\tilde{A}$ to remain invariant, it is then necessary that the potentials shift
under these transformations as
\begin{equation}
  \delta A = \ast \tilde{\rho}, \quad \delta \tilde{A} = \ast \rho
   \label{arho}
\end{equation}
Writing $\rho = \ast \tilde{\lambda}$, $\tilde{\rho} = \ast \lambda$ for
1-form parameters $\lambda, \tilde{\lambda}$, the transformations become
\begin{eqnarray}
  \delta A = \lambda, &  & \delta \tilde{J} = \ast d \lambda  \label{lam}\\
  \delta \tilde{A} = \tilde{\lambda}, &  & \delta J = \ast d \tilde{\lambda} 
  \label{lamt}
\end{eqnarray}
Note that each of the actions that have been discussed depend only on $A$ or only on $\tilde A$ but not both.
The variation of the action (\ref{act}) under (\ref{lam}) is
\begin{equation}
  \delta S = - \int \lambda \wedge \ast j \label{varl}
\end{equation}
If the action is restricted to configurations consistent with the Dirac veto,
then none of the family of Dirac strings $\mathcal{D} (\xi)$ intersect the
world-lines of electric particles and this implies that $\lambda$ vanishes at
any place where $j$ is non-zero. As a result, the Dirac veto ensures that the
variation (\ref{varl}) vanishes and the action is invariant under
(\ref{lamt}).

The alternative action (\ref{acth}) is invariant under (\ref{lam}) but under
(\ref{lamt}) it transforms as
\begin{equation}
  \delta S' = \int F \wedge d \tilde{\lambda}
\end{equation}
Here $F$ is {\emph{defined}} by (\ref{fja}) and so
\begin{equation}
  d F = \ast \tilde{j}
\end{equation}
and as a result
\begin{equation}
  \delta S' = \int \tilde{\lambda} \wedge \ast \tilde{j}
\end{equation}
 This now vanishes as a result of the dual Dirac veto. Similarly, the dual action
(\ref{duact}) is invariant under (\ref{lam}),(\ref{lamt}) provided that
the dual Dirac veto holds. 

Then the Dirac action has $1$-form symmetries corresponding to the symmetry
under changing the
positions of the Dirac strings.
Remarkably, the structure outlined above also appears in the study of
generalised symmetries of Maxwell theory \cite{Gaiotto:2014kfa}; see e.g.\cite{Bhardwaj:2023kri,Brennan:2023mmt,Schafer-Nameki:2023jdn} for reviews and an extensive list of references.
Maxwell theory (without sources) has a
$1$-form symmetry $\delta A = \lambda$ with $d \lambda = 0$. This can be
gauged, i.e.\ promoted to a symmetry for general $\lambda$, by coupling to a
$2$ form gauge field $B$, so that the gauge-invariant field strength is $F = d
A - B$. This agrees with (\ref{fja}) if one takes $B = - \ast \tilde{J}$, so
that the Dirac string current can be interpreted as (the dual of) a   gauge field.
There is a similar story for gauging the dual 1-form symmetry $\delta
\tilde{A} = \tilde{\lambda}$ with gauge field $\tilde{B}$, which can be
identified with $- \ast J$. There is an obstruction to gauging both of these
$1$-form symmetries simultaneously, and this is often expressed by saying
that these symmetries have a mixed anomaly. Then the Dirac veto can be viewed
as a restriction to configurations of the   gauge fields $B,
\tilde{B}$ for which the anomaly vanishes.

This anomaly in the gauged Maxwell theory can be cancelled via anomaly inflow if the theory is
emebedded in a higher dimensional system. If the 4-dimensional spacetime
$\mathcal{M}$ is a boundary of a 5-dimensional space $\mathcal{N}$, then if
$B, \tilde{B}$ are extended to fields on $\mathcal{N}$, then the anomaly
(\ref{varl}) can be cancelled by adding to the action the topological term
\begin{equation}
  \int_{\mathcal{N}} B \wedge d \tilde{B}
\end{equation}
Intriguingly, this suggests that the Dirac veto might be avoided by an
extension of the Dirac theory to a higher dimensional theory.
 
Dirac's theory generalises to a theory of a $p + 1$-form field strength $F$ in
$d$ dimensions in which $j$ is the current density for electrically charged $p
- 1$ branes and $\tilde{j}$ is the current density for magnetically charged $d
- p - 3$ branes \cite{Deser:1997se,Deser:1997mz}. All of the above discussion generalises to general $d, p$,
and in particular the 1-form generalised symmetries described above generalise
to a $p$-form generalised symmetry and a dual $d - p - 2$ form generalised
symmetry. If $p + 1 = d / 2$, the theory can also be extended to allow dyonic
currents \cite{Deser:1997se,Deser:1997mz}.

Following Dirac, the localised currents considered in this paper can be
thought of as differential forms whose components involve delta-functions.
These are examples of  de Rham's generalisation of
differential forms to {\emph{currents}} \cite{Rham}. The singular localised sources
considered here are currents in the sense of \cite{Rham,Griffiths}. The theory of currents
provides the mathematical framework for much of the discussion here.
Such currents have been used previously  to discuss brane sources for $p$-form gauge fields and the corresponding Dirac branes in
\cite{Lechner:2000pn,Cariglia:2004ez,Frey:2019fqz}.

Dirac's theory has played a central role in much subsequent work on monopoles -- see e.g.\
\cite{Blagojevic:1985sh} for an extensive review.
Wu and Yang \cite{Wu:1976qk} provided an elegant alternative to Dirac's formalism in which the
gauge field $A$ is a connection on a $U (1)$ bundle over the submanifold
$\bar{\mathcal{M}}$ of the spacetime $\mathcal{M}$ on which $\tilde{j} = 0$.
Then in each patch $U_{\alpha}$ of a contractible open cover of
$\bar{\mathcal{M}}$ one can introduce a $\tilde{J}_{\alpha} $ and a
$A_{\alpha}$ as above. However, on $\bar{\mathcal{M}}$ one has $\tilde{j} = 0$
and so each $\tilde{J}_{\alpha}$ can be chosen to be zero -- this corresponds
to choosing the Dirac string for $A_{\alpha}$ to lie outside of $U_{\alpha} $
for each $\alpha$. Then $F = d A_{\alpha}$ is a well-defined 2-form on
$\bar{\mathcal{M}}$ and the kinetic term $\int F \wedge \ast F$ is
well-defined. The non-trivial structure is then in the definition of the term
$\int A \wedge \ast j$ in the action as $A$ is now only locally defined. The
definition of this term given in \cite{Wu:1976qk,Alvarez:1984es} leads to a consistency condition in triple
intersections $U_{\alpha} \cap U_{\beta} \cap U_{\gamma}$ that requires
$\frac{1}{2 \pi} F$ to represent an integral cohomology class \cite{Alvarez:1984es}. This results
in an action that is only defined modulo $4 \pi q p$ where $q$ is the electric
charge of any particle and $p$ is the magnetic charge of any monopole and so
leads to the same Dirac quantisation condition as before.

The Wu-Yang approach works well for a heavy magnetic monopole which can be
treated as a defect or (semi-) classical background giving a quantisation of
the electromagnetic field and electrically charged particles in a
 monopole background. However, it is less well adapted to the case of dynamical
magnetic monopoles as the excision of the regions in which $\tilde{j} \neq
0$ can be problematic in practice. In the path integral one would want to
integrate over all possible monopole trajectories, while the monopole wave
function would typically be non-zero over a large region and possibly over the
whole spacetime. As a result, it is not clear what regions of spacetime should be
excised in such a quantum theory. The Dirac approach copes better with dynamical quantum monopoles but still has a number of problems. However, it provides what may be an interesting first step towards a theory electrically charged particles and monopoles in which both are quantum and dynamical.  

The plan of this paper is as follows. In section 2, the generalised symmetry structure of $p$-form gauge theories is reviewed. In section 3, the currents and Dirac strings used in this paper are introduced and some of  the relevant mathematics is discussed. In section 4, Dirac's theory and its field equations are reviewed. In section 5, Dirac's theory is shown to have 1-form symmetries corresponding to deformations of the positions of the Dirac strings. In section 6, it is shown that  the Dirac veto can be regarded as the restriction to 1-form gauge transformations
that are symmetries of the system. In section 7, it is shown that Dirac's action is 
precisely the theory obtained by gauging the 1-form generalised symmetries of Maxwell theory discussed in section 2,  with   2-form gauge fields given in terms of the Dirac string  2-form currents. The Dirac veto corresponds to a restriction to gauge transformations for which there is no anomaly. Section 8 discusses three further aspects.
It discusses the relation between the generalised symmetry of the action and the independence of the field equations from the positions of the Dirac strings, the generalisation from symmetries of the classical action to the larger set of symmetries of the quantum theory, and the possibility of generalising the Dirac theory to a more general class of theories.
Section 9
discusses the possibility of cancelling the anomaly by embedding the theory in higher dimensions, with the 4-dimensional spacetime arising either as a boundary of a 5-dimensional theory, or as a 3-brane in a higher-dimensional theory, such as IIB string theory.
In section 9 the discussion of Maxwell theory is generalised to a $p$-form gauge field in $d$ dimensions coupled to charged branes, while in section 10 it is generalised to non-linear theories such as Born-Infeld theory. In section 11, a global treatment is given and related to the Wu-Yang approach and in section 12 there is further discussion of the theories analysed in this paper, together with possible generalisations.

\section{Generalised Symmetries and $p$-Form Gauge Theories}

\subsection{Generalised Symmetries}\label{GenSym}

Generalised symmetries were introduced in \cite{Gaiotto:2014kfa}; see e.g.\ \cite{Bhardwaj:2023kri,Brennan:2023mmt,Schafer-Nameki:2023jdn} for reviews of the field.
In this section, some features of the generalised symmetries of $p$-form gauge theories will be reviewed.

A $(p + 1)$-form field strength $F$ in $d$ dimensional Minkowski space
satisfies the equations\footnote{In $d$ dimensions with Lorentzian signature,
for an $r$-form $\omega$ the Hodge duality operator satisfies $\ast^2 \omega =
- (- 1)^{r (d - r)} \omega$ and $d^{\dagger} \omega = (- 1)^{d (r + 1)} \ast d
\ast \omega$.}
\begin{equation}
  dF = \ast \tilde{j}  \qquad d \ast F = \ast j \label{Ffida}
\end{equation}
where $j$ is a $p$-form electric current and $\tilde{j}$ is a $\tilde{p}$-form
magnetic current with $\tilde{p} = d - p - 2$; these currents are required to
be conserved, $d \ast j = 0$, $d \ast \tilde{j} = 0$. In addition, $F$ and
$\ast F$ can be viewed as $(p + 1)$-form and $(\tilde{p} + 1)$-form currents
that are conserved in regions where the corresponding currents $j, \tilde{j}
$ vanish. Explicitly, $\ast F$ is conserved, $d^{\dag} \ast F$=0, in regions in
which $\tilde{j} = 0$ while $F$ is conserved, $d^{\dag} F$=0, in regions in
which $j = 0$.

If there are no magnetic sources, i.e.\ if $\tilde{j} = 0$, then there is a
local $p$-form potential $A$ with $F = dA$ and a gauge symmetry $\delta A = d
\sigma$ with $(p - 1)$-form parameter $\sigma$. The action is
\begin{equation}
  S = \frac{1}{2} \int F \wedge \ast F + A \wedge \ast j
\end{equation}
If $j = 0$, then the action is invariant under the $p$-form transformation
\begin{equation}
  \delta A = \lambda
\end{equation}
where $\lambda$ is a closed $p$-form. If $\lambda$ is exact, $\lambda = d
\sigma$, then this is a gauge transformation with parameter $\sigma$, so the
new symmetries correspond to closed forms modulo exact forms, i.e.\ to the
cohomology classes $[\lambda] .$

Similarly, if there are no electric sources, i.e.\ if $j = 0$, then there is a
local $\tilde{p}$-form potential $\tilde{A}$ with $\ast F = d \tilde{A}$ and a
gauge symmetry $\delta \tilde{A} = d \tilde{\sigma}$ with $\tilde{p} - 1$-form
parameter $\tilde{\sigma}$. If $\tilde{j} = 0$, then the dual formulation is
invariant under the  $\tilde p$-form transformation
\begin{equation}
  \delta \tilde{A} = \tilde{\lambda}
\end{equation}
where $\tilde{\lambda}$ is a closed $\tilde{p}$-form, and the new symmetries
correspond to the cohomology classes $[\tilde{\lambda}] $. As the parameters
$\lambda, \tilde{\lambda}$ are constrained, these can be regarded as
defining global symmetries. 

Consider now the free  theory without sources, $j = \tilde{j} = 0$, which has both
of these symmetries. Gauging the $\lambda$ symmetry  promotes
it to a symmetry for  parameters that are unconstrained 1-form fields $\lambda (x)$ by coupling to a $p +
1$ form gauge field $B$. Similarly,    gauging the $\tilde{\lambda}$ symmetry   promotes it to a symmetry for unconstrained position-dependent parameters
$\tilde{\lambda}(x)$ by coupling to a $\tilde{p} + 1$ form gauge field
$\tilde{B}$. The gauge fields transform as
\begin{equation}
  \delta B = d \lambda, \quad \delta \tilde{B} = d \widetilde{\lambda}
\end{equation}
so that
\begin{equation}
  F = d A - B, \quad \tilde{F} = d \tilde{A} - \tilde{B} \label{FAB}
\end{equation}
are gauge-invariant.

Consider first the formulation in terms of $A$ (without sources). Then
\begin{equation}
  S = \frac{1}{2} \int F \wedge \ast F
\end{equation}
with $F = d A - B$ is invariant under the $\lambda$-transformations. The
$\tilde{\lambda}$ transformations do not act on $A$;    here  gauging means
coupling the gauge field $\tilde{B} $ to the current $\ast F$. The action
\begin{equation}
  \int d A \wedge \ast d A + \tilde{B} \wedge d A \label{AdA}
\end{equation}
is invariant under the $\tilde{\lambda}$-transformations.

There is an obstruction to gauging both symmetries. For example,
\begin{equation}
  S = \int \frac{1}{2} F \wedge \ast F + \tilde{B} \wedge d A \label{actaa}
\end{equation}
is invariant under the $\tilde{\lambda}$ transformations but under the
$\lambda$ transformations
\begin{equation}
  \delta S = \int d \lambda \wedge \tilde{B} \label{anomaly}
\end{equation}
This can be cancelled by adding a counterterm so that $d A$ is replaced with
$F = d A - B$ in (\ref{AdA}) so that
\begin{equation}
  S = \int \frac{1}{2} F \wedge \ast F + \tilde{B} \wedge F  \label{actbb}
\end{equation}
This is now invariant under the $\lambda$ transformations but under the
$\tilde{\lambda}$ transformations changes by
\begin{equation}
  \delta S = - \int d \tilde{\lambda} \wedge B \label{anomya}
\end{equation}
This obstruction to gauging in the classical theory is referred to as a mixed
anomaly in the global symmetries of the ungauged theory.

For a given background gauge field $\tilde{B} $, the anomalous variation (\ref{anomaly})  will vanish for those $\lambda$ with
\begin{equation}
  \int  \lambda \wedge \tilde{H} =0 \label{anomalyaa}
\end{equation}
with $\tilde{H}=d \tilde{B}$, so that there is a subgroup of the 1-form symmetry that is non-anomalous. In particular, this vanishes for exact 1-forms $\lambda$, corresponding to standard gauge symmetries.
In the quantum theory, it is sufficient that the action be invariant modulo $2\pi \hbar$.
Under the $\lambda $ transformation,  (\ref{anomaly}) then gives the condition for quantum invariance to be
\begin{equation}
  \int  \lambda \wedge \tilde{H} \in 2\pi \hbar \mathbb{Z} \label{anomalybb}
\end{equation}
so that there can be a slightly larger remnant of the 1-form symmetry.
Similar remarks apply to (\ref{anomya}).

The anomaly can be cancelled if the $d$-dimensional theory is coupled to a
topological theory in $d + 1$ dimensions. The gauge fields $B, \tilde{B}$ are
extended into a space $\mathcal{N}$ in $d + 1$ dimensions whose boundary is the
$d$-dimensional spacetime $\mathcal{M}$, $\partial \mathcal{N} = \mathcal{M}$. Then adding a term
\begin{equation}
  \int_{\mathcal{N}} B \wedge d \tilde{B}
\end{equation}
to the action cancels the anomaly (\ref{anomaly}). This anomaly structure has
many applications and implications; see e.g.\ \cite{Bhardwaj:2023kri,Brennan:2023mmt,Schafer-Nameki:2023jdn}  for further discussion.
Note that in this section the gauge fields $B$,$\tilde B$ have been treated as background fields which are not integrated over.

\subsection{Charges and Currents}

The charge generating the $\lambda $ symmetry can be found following \cite{Hull:2023iny}.
Consider the action $\int F\wedge *F$ with $F=dA$.
For a {\emph {fixed}} closed 1-form $\lambda$, the theory is invariant under 
\begin{equation}
\delta A= \alpha \lambda
\end{equation}
for any real number $\alpha$. This can be thought of as global 0-form symmetry with parameter $\alpha$.
(For the gauged theory, the gauge transformation with parameter $\alpha \lambda$  is an invariance of the background gauge field, $\delta B=0$.)
If $\alpha $ is taken to be a function, the variation of the action is proportional to $\int d \alpha \wedge *j_N$
where $j [\lambda]$ is
the   1-form Noether current given by
\begin{equation}
  \ast j [\lambda] =  \lambda \wedge \ast F
\end{equation}
with corresponding Noether charge
\begin{equation}
  Q [\lambda] = \int_{\Sigma} \lambda \wedge \ast F
\end{equation}
This charge then generates the
$\lambda$-transformation
\begin{equation}
  [Q [\lambda], A] = \lambda
\end{equation}
so that acting on $A$ gives the transformation $ \delta A = \lambda
$. This is a Poisson bracket
relation: for the quantum commutator there is an extra factor of $i$.

This is easily checked in the canonical formalism for the ungauged theory
(with $B = 0$). Consider e.g.\ the case in which $A$ is a 1-form with time
component $A_0$ and spatial components $A_i$ with $i = 1, \ldots, d - 1$ and
take $\Sigma$ to be a spatial hypersurface with constant $x^0$. The momentum
conjugate to $A_i$ is $\pi_i = F_{0 i}$ while $A_0$ is a lagrange multiplier
for the Gauss constraint.

Then (with $n = d - 1$)
\begin{equation}
  Q [\lambda] = \int d^n x \, \lambda^i \pi_i
\end{equation}
and the canonical Poisson brackets then give
\begin{equation}
  [Q [\lambda], A_i] = \lambda_i
\end{equation}
(For the quantum commutators, there is an extra factor of $i$ in this
relation.) A similar construction was given for linear gravity in \cite{Hull:2024bcl}.

Similarly, for the formulation in terms of $\tilde{A}$, for a closed
$\tilde{\lambda}$ the Noether current for the $\tilde{\lambda}$ symmetry is
\begin{equation}
  \ast j [\tilde{\lambda}] =  \tilde{\lambda} \wedge F
\end{equation}
with Noether charge
\begin{equation}
  Q [\tilde{\lambda}] = \int_{\Sigma} \tilde{\lambda} \wedge F
\end{equation}
that generates the $\tilde{\lambda}$-transformation
\begin{equation}
  [Q [\tilde{\lambda}], \tilde{A}] = \tilde{\lambda}
\end{equation}
Again, this is easily checked using the canonical formalism for the theory
formulated in terms of $\tilde{A}$.

\section{Currents and Dirac Strings}

Consider a particle of electric charge $q$ moving in a $d$-dimensional spacetime $\mathcal{M}$ with metric $g_{\mu\nu}$.
The particle's world-line is a curve
$\mathcal{C}$ in $\mathcal{M}$, parameterised by some $\tau$, and which is specified by $x^{\mu} = X^{\mu}
(\tau)$ for some function $X^{\mu} (\tau)$. Its coupling to the 1-form gauge
field $A$ is given by
\begin{equation}
  q \int_{\mathcal{C}} A = q \int d \tau A_{\mu} (X (\tau)) \frac{d X^{\mu}}{d
  \tau} \label{qAis}
\end{equation}
This can be written as
\[ \int_{\mathcal{M}} A \wedge \ast j \label{qA} \]
where the electric current is localised on the world-line:
\begin{equation}
  j^{\mu} (x) = q {\int_{\mathcal{C}}}  d \tau \frac{d X^{\mu}}{d \tau} \delta
  (x - X (\tau)) \label{jlin}
\end{equation}
For a curve with end-points from $X (\tau_1)$ to $X (\tau_2)$, the current is
\quad
\begin{equation}
  j^{\mu} (x) = q {\int_{\tau_1}^{\tau_2}}  d \tau \frac{d X^{\mu}}{d \tau}
  \delta (x - X (\tau)) \label{jlin2}
\end{equation}
and satisfies
\begin{equation}
  \partial_{\mu} j^{\mu} (x) = q [\delta (x - X (\tau_2)) - \delta (x - X
  (\tau_1))]
\end{equation}
Thus for an infinite world-line (with $X^0 (\tau_2) = \infty$ and $X^0
(\tau_1) = - \infty$) or for a closed curve (with $X^{\mu} (\tau_2) = X^{\mu}
(\tau_1)$) the current is conserved, and attention will be restricted to these
cases here. For a finite world-line, the delta-functions represent the
creation of a particle at $X (\tau_1)$ and its destruction at $X (\tau_2)$.

Similarly, a $p$-form gauge field $A$ couples to a charged $p - 1$ brane. If
the brane world-volume is a $p$-dimensional submanifold $\mathcal{N} \subset
\mathcal{M}$, then the coupling can be written as
\begin{equation}
  q \int_{\mathcal{N}} A = \int_{\mathcal{M}} A \wedge \ast j \label{qA2}
\end{equation}
where
\begin{equation}
  j^{\mu_1 \ldots \mu_{q - 1}} = 
  \frac q {(q-1!)}
    \int d^{q - 1} \sigma  \,
  \varepsilon^{a_1 a_2 \ldots a_{q - 1}}  \frac{\partial X^{\mu_1}}{\partial
  \sigma ^{a_1}} \frac{\partial X^{\mu_2}}{\partial \sigma ^{a_2}} \ldots
  \frac{\partial X^{\mu_{q - 1}}}{\partial \sigma ^{a_{q - 1}}} \delta (x - X
  (\sigma )) \label{jcuris}
\end{equation}
Here $\sigma^a$ ($a = 0, 1, \ldots, p - 1$) are world-volume coordinates and
the submanifold $\mathcal{N} \subset \mathcal{M}$ is specified by $x^{\mu} =
X^{\mu} (\sigma^a)$ for some functions $X^{\mu} (\sigma^a)$, while $\varepsilon^{a_1 a_2 \ldots a_{q - 1}}$ is an antisymmetric tensor density.
 Again, this is
conserved ($d^\dag j=0$) for branes extending from the infinite past to the infinite future,
or for closed branes.

It is useful to introduce the $p$-form delta-function associated with a
$p$-dimensional submanifold $\mathcal{N} \subset \mathcal{M}$. This is a
$p$-form such that for any $p$-form $\omega$ on $\mathcal{M}$
\begin{equation}
  \int_{\mathcal{N}} \omega = \int_{\mathcal{M}} \omega \wedge \ast
  \delta_{\mathcal{N}}
   \label{PD}
\end{equation}
The $d-p$ form $ \ast
  \delta_{\mathcal{N}}
$
is sometimes referred to as the Poincar\' e dual of $\mathcal{N} \subset \mathcal{M}$.
If $\mathcal{N} \subset \mathcal{M}$ is specified by $x^{\mu} = X^{\mu}
(\sigma^a)$, then $ \ast
  \delta_{\mathcal{N}}
$
can be written explicitly as
\begin{equation}
    \begin{aligned}
 \ast \delta_{\mathcal{N}} (x) =& \frac 1 {p!(d-p!)} \int d^p \sigma  \, \varepsilon^{a_1 a_2
  \ldots a_p}  \frac{\partial X^{\mu_1}}{\partial \sigma ^{a_1}}
  \frac{\partial X^{\mu_2}}{\partial \sigma ^{a_2}} \ldots \frac{\partial
  X^{\mu_p}}{\partial \sigma ^{a_p}} \\
  &
  \times
  \epsilon_{\mu_1 \mu_2 \dots \mu_p  \nu_1\dots \nu_{d-p}} dx^{ \nu_1}\wedge \dots dx^{ \nu_{d-p}}
  \delta (x - X (\sigma )) \label{poidual}
\end{aligned}
\end{equation}
so that
the delta-function form can be written explicitly as
\begin{equation}
  \delta_{\mathcal{N}} (x) =
  \frac 1 {(p!)^2}
   \int d^p \sigma  \, \varepsilon^{a_1 a_2
  \ldots a_p}  \frac{\partial X^{\mu_1}}{\partial \sigma ^{a_1}}
  \frac{\partial X^{\mu_2}}{\partial \sigma ^{a_2}} \ldots \frac{\partial
  X^{\mu_p}}{\partial \sigma ^{a_p}} \delta (x - X (\sigma )) dx_{ \mu_1}\wedge \dots dx_{ \mu_{p}}
  \label{delfis}
\end{equation}
where $dx_\mu = g_{\mu\nu}dx^\nu$.
Then (\ref{jlin}) can be written as
\begin{equation}
  j = q \delta_{\mathcal{C}}
\end{equation}
and (\ref{jcuris}) can be written as
\begin{equation}
  j = q \delta_{\mathcal{N}}
\end{equation}

If a $q$-dimensional submanifold $\mathcal{P}$ has boundary $\partial
\mathcal{P}$, then for a $q - 1$ form $\sigma$
\begin{equation}
  \int_{\partial \mathcal{P}} \sigma = \int_{\mathcal{P}} d \sigma 
 \end{equation}
This can then be written as
\begin{equation}
  \int_{\mathcal{M}} \sigma \wedge \ast \delta_{\partial \mathcal{P}} =
  \int_{\mathcal{M}} d \sigma \wedge \ast \delta_{\mathcal{P}} =
  \int_{\mathcal{M}} \sigma \wedge \ast d^{\dag} \delta_{\mathcal{P}} +
  \int_{\mathcal{M}} d (\sigma \wedge \ast \delta_{\mathcal{P}})
  \label{eheadscf}
\end{equation}
If $\mathcal{P}$ has no intersection with the boundary of $\mathcal{M}$ (or if
$\mathcal{M}$ has no boundary), then $\delta_{\mathcal{P}}$ will be zero on
the boundary and the last term in (\ref{eheadscf}) vanishes. Then
\begin{equation}
  \delta_{\partial \mathcal{P}} = d^{\dag} \delta_{\mathcal{P}}
  \label{ddagdel}
\end{equation}
as can be checked explicitly using (\ref{delfis}). This will be used below for $\partial
\mathcal{P}={\mathcal{N}}$.

As discussed in the introduction, singular forms such as
$\delta_{\mathcal{N}}$ are examples of what mathematicians call
{\emph{currents}}, as defined in \cite{Rham,Griffiths}. Currents are maps from the space of
smooth forms to the real numbers, and in this case the map for a given
$\mathcal{N}$ is given by (\ref{PD}). This is reminiscent of Poincar\' e duality, where
for a {\emph{closed}} $p$-form $\omega$, the integral $\int_{\mathcal{N}}
\omega$ over a cycle $\mathcal{N}$ depends only on the homology class of
$\mathcal{N}$ and can be written as

\begin{equation}
  \int_{\mathcal{N}} \omega = \int_{\mathcal{M}} \omega \wedge \ast
  \alpha_{\mathcal{N}}
\end{equation}
where $\alpha_{\mathcal{N}}$ is a closed $p$-form and the result depends only
on the cohomology class of $\alpha_{\mathcal{N}}$. The cohomology class
$[\alpha_{\mathcal{N}}]$ is the Poincar\' e dual of the homology class of
$\mathcal{N}$. The result (\ref{PD}) here is a generalisation of this in which the
$p$-form $\omega$ need not be closed and $\mathcal{N}$ can be any $p$-chain.
Then this defines a duality between the current $\delta_{\mathcal{N}}$ and the
chain $\mathcal{N}$. In the physics literature,  $\delta_{\mathcal{N}}$ is sometimes referred to as
the Poincar\' e dual of $\mathcal{N}$.

For the case of a 1-form potential $A$ with $\mathcal{C}$ a closed curve, the
exponential of the coupling (\ref{qAis}) (multiplied by $i$) defines a
Wilson line operator. If $\mathcal{D}$ is a 2-surface with boundary $\mathcal{C}$, then
the coupling (\ref{qAis}) can be written as
\begin{equation}
  q \int_{\mathcal{C}} A = q \int_{\mathcal{D}} F \label{afcd}
\end{equation}
which can be re-expressed as
\begin{equation}
  \int_{\mathcal{M}} A \wedge \ast j = \int_{\mathcal{M}} F \wedge \ast J
  \label{ajfjis}
\end{equation}
where
\begin{equation}
  J = q \delta_{\mathcal{D}}
\end{equation}
and satisfies $d^{\dag} J = j$ as a result of (\ref{ddagdel}). However, $q
\int_{\mathcal{D}} F$ depends on the choice of surface with boundary
$\mathcal{C}$. For two surfaces $\mathcal{D}, \mathcal{D}'$ with boundary
$\mathcal{C}$,
\begin{equation}
  q \int_{\mathcal{D}'} F - q \int_{\mathcal{D}} F = q \int_{\mathcal{E}} F
\end{equation}
where $\mathcal{E}=\mathcal{D}  \cup \mathcal{D}'$ is the closed surface given
by combining $\mathcal{D}, \mathcal{D}'$ with opposite orientations. Note that
$p = \int_{\mathcal{E}} F$ is the magnetic charge contained in $\mathcal{E}$
(which is $2 \pi$ times an integer if $F$ is conventionally normalised). Then
the Wilson line
\begin{equation}
  W (\mathcal{C}) = e^{\frac{i}{\hbar} q \int_{\mathcal{D}} F}
\end{equation}
changes by a phase
\[ e^{\frac{i}{\hbar} q p} \]
on changing from $\mathcal{D}'$ to $\mathcal{D}$ and so is well-defined
provided that the charges satisfy the Dirac quantisation condition
\begin{equation}
  p q = 2 \pi n \hbar, \quad n \in \mathbb{Z}
  \label{Diracquant}
\end{equation}
for some integer $n$.

In general, $A$ need not be globally defined. Given an open cover of
$\mathcal{M}$ by contractible open sets $U_{\alpha}$, there will be a 1-form
$A_{\alpha}$ in each $U_{\alpha}$ with $F = d A_{\alpha}$ and in each overlap
\begin{equation}
  A_{\alpha} - A_{\beta} = d \lambda_{\alpha \beta} \quad \tmop{in} \quad U_{\alpha}
  \cap U_{\beta}
\end{equation}
If the curve $\mathcal{C}$ is contained in one patch $U_{\alpha}$, then $q
\int_{\mathcal{C}} A$ is well-defined. If the curve passes through several
patches, however, $q \int_{\mathcal{C}} A$ is not well-defined but $q
\int_{\mathcal{D}} F$ is well-defined and can be taken as the definition of
the coupling.

Now consider the case in which $\mathcal{C}$ is not closed but is the
world-line of a charged particle. For each $\tau$,    a Dirac string is introduced that
goes from the particle to infinity and which is specified by functions $Y^{\mu} (\tau,
\sigma)$ with $Y^{\mu} (\tau, 0) = X^{\mu} (\tau)$ so that the string
world-sheet $\mathcal{D}$ is specified by $x^{\mu} = Y^{\mu} (\tau, \sigma)$.
The boundary of $\mathcal{D}$ is $\mathcal{C} \cup \mathcal{G}$ where
\ensuremath{\mathcal{G}} is the part of the boundary at infinity.

For example, for a particle at the origin, $\mathcal{C}$ is the line $(t,
\vec{x})$ with spatial coordinates $\vec{x} = 0$ and the Dirac string can be
taken to be along the $x$-axis from $x = 0$ to some $x = R$ in the limit $R
\rightarrow \infty$. The time $t$ can be taken in the range $t \in (- T, T)$
for some large $T$ which will also be taken to infinity. The Dirac string
world-sheet $\mathcal{D}$ is then a large rectangle of sides $R$ and $2 T$, in
the limit $R \rightarrow \infty$, $T \rightarrow \infty$. The boundary of the
rectangle consists of $\mathcal{C}$ which is the line $(t, \vec{x})$ with
spatial coordinates $\vec{x} = 0$ and the time taken to be in the range $t \in
(- T, T)$ 
together with the curve $\mathcal{G}$ which
consists of the line from $(- T, 0, 0, 0)$ to $(- T, R, 0, 0)$, the line from
$(- T, R, 0, 0)$ to $(T, R, 0, 0)$ and the line from $(T, R, 0, 0)$ to $(T, 0,
0, 0)$. The limit $R \rightarrow \infty$, $T \rightarrow \infty$ is then taken.

For such a world-line, (\ref{afcd}) becomes
\begin{equation}
  q \int_{\mathcal{C}} A = q \int_{\mathcal{D}} F - q \int_{\mathcal{G}} A
\end{equation}
and (\ref{afcd}) only holds with suitable boundary conditions, e.g.\ if $A = 0$
on $\mathcal{G}$, in which case one has (\ref{ajfjis}). Note that the actions
$q \int_{\mathcal{C}} A$ and $q \int_{\mathcal{D}} F$ give the same field
equations from variations that vanish on $\mathcal{G}$.

Some of the formulae in this paper involve products of currents. If these are
delta-function currents, such products can be ill-defined. As in \cite{Dirac:1948um},  it
will be supposed here that the delta functions are smeared to some smooth functions
$\delta_{\varepsilon} (x)$ which integrate to one ($\int d x$
$\delta_{\varepsilon} (x)$=1) and have compact support ($\delta_{\varepsilon}
(x) = 0$ for $| x | > \varepsilon$) which tend to the Dirac delta-function as
$\varepsilon \rightarrow 0$. The $\varepsilon$ dependence will not be shown
explicitly and the limit $\varepsilon \rightarrow 0$ will usually be assumed
when the final result is non-singular and unambiguous.

\section{Dirac's Theory and Dirac Strings}\label{DiracTh}

Consider a set of $N$ particles in 4 dimensions labelled by $i = 1, \ldots, N$
with masses $m_i$, electric charges $q_i$ and magnetic charges $p_i$ moving
along worldlines $\mathcal{C}_i$ given by $x^{\mu} = X_i^{\mu} (\tau_i)$ for
some functions $X^{\mu }_i (\tau_i)$. Then the electric and magnetic currents
are the 1-forms
\begin{equation}
  j (x) = \sum_{i = 1}^N q_i \delta_{\mathcal{C}_i} (x), \quad \tilde{j} (x) =
  \sum_{i = 1}^N p_i \delta_{\mathcal{C}_i} (x) \label{jiss}
\end{equation}
Dirac considered the special case in which each particle had either electric charge or
magnetic charge but not both, so that for each $i$ either $q_i = 0$ or $p_i =
0$ \cite{Dirac:1948um}. The generalisation to allow each particle to be dyonic was considered
in \cite{Deser:1997se,Deser:1997mz} and will be analysed here; the discussion is similar to Dirac's case but there are some extra
subtleties, as will be seen below.

For each particle, a Dirac string is located on a curve from the particle to
infinity and its evolution gives a world-sheet with a boundary on the particle
world-line. For a worldline $\mathcal{C}_i $ the Dirac string has world-sheet
on a 2-surface $\mathcal{D}_i$ given by $x^{\mu} = Y_i^{\mu} (\tau_i,
\sigma_i)$ where $\sigma_i \geqslant 0$ and the functions $ Y_i^{\mu} (\tau_i,
\sigma_i)$ satisfy the boundary condition $Y_i^{\mu} (\tau_i, 0) = X_i^{\mu}
(\tau_i)$. For fixed $\tau_i$, the Dirac string $x^{\mu} = Y_i^{\mu} (\tau_i,
\sigma_i)$ is a curve from the particle at $X_i^{\mu} (\tau_i)$ to infinity.
Then
\begin{equation}
  \tilde{j} = d^{\dag} \tilde{J}
  \label{wetws}
\end{equation}
where the secondary current is
\begin{equation}
  \tilde{J} = \sum_i p_i \delta_{\mathcal{D}_i}
  \label{Jtiis}
\end{equation}
and is localised on the Dirac string
and (\ref{wetws}) follows from (\ref{ddagdel}).
The equation
\begin{equation}
  d F = \ast \tilde{j}
\end{equation}
is then solved by taking
\begin{equation}
  F = d A + \ast \tilde{J} \label{FAJ}
\end{equation}
Choosing the parameter $\tau_i$ to be the proper time on the curve
$\mathcal{C}_i $ for each $i$, the remaining field equations are then
\begin{equation}
  d^{\dag} F = j \label{ElMax}
\end{equation}
and
\begin{equation}
  m_i \frac{d^2 X^{\mu}_i}{d \tau_i^2} = (q_i F^{\mu \nu} + p_i \ast F^{\mu
  \nu}) \eta_{\nu \rho} \frac{d X_i^{\rho}}{d \tau_i} \label{Xeq}
\end{equation}

Dirac's   action  for these field equations is $I = I_1 + I_2 + I_3$
where
\begin{equation}
  I_1 = \sum_i m_i \int_{\mathcal{C}_i} d \tau_i \label{act1}
\end{equation}
\begin{equation}
  I_2 = \int \frac{1}{2} F \wedge \ast F \label{Facti}
\end{equation}
\begin{equation}
  I_3 = \int d^4 x A_{\mu} j^{\mu} = \sum_i q_i \int_{\mathcal{C}_i} d \tau_i
  A_{\mu} (X_i (\tau_i)) \frac{d X^{\mu}_i}{d \tau_i} \label{act3}
\end{equation}
where $F$ is given by (\ref{FAJ}).
The action is regarded as a functional of the vector potential $A^{\mu}
(x)$, the particle worldlines $X^{\mu }_i (\tau_i)$ and the Dirac string
world-sheets $Y_i^{\mu} (\tau_i, \sigma_i)$. 
Note that the action depends only on $\tilde J$ and not $J$, so that the Dirac strings only enter for those particles with magnetic charge, for which $p_i\ne 0$. There is a dual formulation which depends only on the Dirac strings for particles with electric charge. It is useful to present results for the general case in which all particles can be dyonic, while noting the simplifications that arise for the cases in which all particles have either electric or magnetic charge, but not both.

The variations of these actions
are then
\begin{equation}
  \delta I_1 = \sum_i m_i \int_{\mathcal{C}_i} d \tau_i  \frac{d^2
  X^{\mu}_i}{d \tau_i^2} \delta X^{\nu}_i \eta_{\mu \nu}
\end{equation}
\begin{equation}
  \delta I_2 = \int d^{\dag} F \wedge \ast \delta A + F \wedge \delta
  \tilde{J}
\end{equation}
\begin{equation}
  \delta I_3 = \sum_i q_i \int_{\mathcal{C}_i} d \tau_i [(\partial_{\mu}
  A_{\nu} - \partial_{\nu} A_{\mu}) \delta X^{\nu}_i + \delta A_{\mu} (X_i
  (\tau_i))] \frac{d X^{\mu}_i}{d \tau_i} \label{deli3}
\end{equation}
Now
\begin{equation}
  \int_{\mathcal{M}} F \wedge \tilde{J} = \sum_i p_i \int_{\mathcal{D}_i} d^2
  \sigma_i \epsilon^{a b} (\ast F)_{\mu \nu} (Y^{\rho}_i (\sigma^c_i)) 
  \frac{\partial Y^{\mu}_i}{\partial \sigma_i^a} \frac{\partial
  Y^{\nu}_i}{\partial \sigma_i^b}
\end{equation}
where the world-sheet coordinates are written $\sigma^a_i = (\tau_i,
\sigma_i)$ with $a, b = 0, 1$. Varying $Y^{\rho}_i (\sigma^c_i)$ gives
\begin{equation}
  \int F \wedge \delta \tilde{J} = \sum_i p_i \int_{\mathcal{D}_i} d^2
  \sigma_i 
  \left[
  \epsilon^{a b} \partial_{[\rho } (\ast F)_{\mu 
  \nu]}  \frac{\partial Y^{\mu}_i}{\partial \sigma_i^a} \frac{\partial
  Y^{\nu}_i}{\partial \sigma_i^b} \delta Y^{\rho}_i + \partial_a \left(
  \epsilon^{a b} (\ast F)_{\mu \nu} \frac{\partial Y^{\nu}_i}{\partial
  \sigma_i^b} \delta Y_i^{\mu} \right) \label{FJvar}
  \right]
\end{equation}
The second term in the parenthesis is a total derivative and can be rewritten as the  integral over the boundary $\partial
\mathcal{D}_i$ of $\mathcal{D}_i$, part of which is the world-line
$\mathcal{C}_i$. Then $\partial \mathcal{D}_i = \mathcal{C}_i \cup
\mathcal{G}_i$ where the remainder of the boundary is a curve $\mathcal{G}_i$
at infinity. With suitable boundary conditions (e.g.\  $\delta Y_i^{\mu} = 0$ on
$\mathcal{G}_i$), the second term reduces to a sum of integrals over the
world-lines $\mathcal{C}_i$
\begin{equation}
  \sum_i p_i \int_{\mathcal{D}_i} d^2 \sigma_i \partial_a \left( \epsilon^{a
  b} (\ast F)_{\mu \nu} \frac{\partial Y^{\nu}_i}{\partial \sigma_i^b} \delta
  Y_i^{\mu} \right) = \sum_i p_i \int_{\mathcal{C}_i } d \tau_i  (\ast F)_{\mu
  \nu} \frac{d X^{\mu}_i}{d \tau_i} \delta X^{\nu} \label{pF}
\end{equation}
Then the variation $\delta A$ in $I_2 + I_3$ gives the field equation
(\ref{ElMax}) as required. The variation $\delta X$ from $I_1 + I_2 + I_3$
gives, using (\ref{FAJ}),(\ref{FJvar}) and (\ref{pF}), the required field
equation (\ref{Xeq}) provided
\begin{equation}
  \sum_i q_i \int_{\mathcal{C}_i} d \tau_i [(\ast \tilde{J})_{\mu \nu}
  (X^{\mu}_i (\tau_i))] \delta X^{\nu}_i \frac{d X^{\mu}_i}{d \tau_i} = 0
  \label{Dircon}
\end{equation}
which can be rewritten as
\begin{equation}
  \tilde{J} \wedge j = 0 \label{veto}
\end{equation}
(This is the condition that allows $d A$ to be replaced by $F$ in
(\ref{deli3}).) Note that in (\ref{Dircon}) $(\ast \tilde{J})_{\mu \nu}
(X^{\mu}_i (\tau_i))$ is the 2-form $\ast J$ at the point $X^{\mu}_i (\tau_i)$
on $\mathcal{C}_i$, so (\ref{Dircon}) requires (\ref{veto}) to hold on
$\mathcal{C}_i$. However $j$ is only non-zero on $\mathcal{C}_i$, so as a
result (\ref{veto}) is true everywhere.

For Dirac's case, each particle either has electric charge or magnetic charge but not both.
 Let $i =
(a, \alpha)$ with
$a$ labelling the electric charges (so that $p_a = 0$) and $\alpha$ labelling the magnetic ones (so that  $q_{\alpha} = 0$).
 Then only the Dirac strings
$\mathcal{D}_{\alpha}$ for the magnetically charged particles enter the
formalism. As a result, (\ref{Dircon}) becomes
\begin{equation}
  \sum_a q_a \int_{\mathcal{C}_a} d \tau_a (\ast \tilde{J})_{\mu \nu}
  (X^{\mu}_a (\tau_a)) \delta X^{\nu}_a \frac{d X^{\mu}_a}{d \tau_a} = 0
  \label{ertr}
\end{equation}
and this will hold if the position of each magnetic Dirac string world-sheet
$\mathcal{D}_{\alpha}$ is chosen so that no electrically charged particle
world-line $\mathcal{C}_a$ intersects it, i.e.
\begin{equation}
  \mathcal{D}_{\alpha} \cap \mathcal{C}_a = 0
  \label{dirvetcase}
\end{equation}
for all $a, \alpha$. This will ensure that $(\ast \tilde{J})_{\mu \nu}
(X^{\mu}_a (\tau_a)) = 0$ so that (\ref{ertr}) vanishes. This condition is the
Dirac veto.

Consider now the general case of dyonic particles. The expression (\ref{veto})
is a sum of terms with contributions from particle worldlines $\mathcal{C}_j$
and Dirac string world-sheets $\mathcal{D}_i$, with a summation over all $i,
j$. Writing the 2-form current $J$ restricted to $\mathcal{D}_i$ as
$\tilde{J}_i = p_i \delta_{\mathcal{D}_i}$ and the 1-form current $j$
restricted to $\mathcal{C}_i$ as $j_i (x) = q_i \delta_{\mathcal{C}_i} (x)$,
(\ref{veto}) is
\begin{equation}
  \sum_{i, j} \tilde{J}_j \wedge j_i = 0
\end{equation}
The terms with $j \neq i$ will vanish,
\begin{equation}
  \sum_{i \neq j} \tilde{J}_j \wedge j_i = 0
\end{equation}
if the position of each string world-sheet $\mathcal{D}_i$ is chosen so that
no particle world-line $\mathcal{C}_j$ for $j \neq i$ intersects it, i.e.
\begin{equation}
  \mathcal{D}_i \cap \mathcal{C}_j = 0 \qquad \tmop{for} j \neq i
\end{equation}
This formulates the generalisation of the Dirac veto given in \cite{Deser:1997se,Deser:1997mz}. For the terms with
$i = j$, $\mathcal{C}_i$ is tangent to $\mathcal{D}_i$ so that the terms
$\tilde{J}_j \wedge j_i$ with $i = j$ formally vanish. This expression involves the
product of two delta functions and it will be assumed that the delta-functions
are regularised in such a way that $\tilde{J}_i \wedge j_i = 0$ for each $i$, as  in \cite{Deser:1997se,Deser:1997mz}.

The remaining field equation is that from varying the position of the Dirac
string $Y$, giving the first term  in the parenthesis in (\ref{FJvar}). However, using
(\ref{ElMax}), this can be rewritten as
\begin{equation}
  \sum_i p_i \int_{\mathcal{D}_i} d^2 \sigma_i \epsilon^{a b} (\ast j)_{\mu
  \nu \rho}  \frac{\partial Y^{\mu}_i}{\partial \sigma_i^a} \frac{\partial
  Y^{\nu}_i}{\partial \sigma_i^b} \delta Y^{\rho}_i =  \sum_i \int_{\mathcal{M}} \ast
  [(\ast j)_{\mu \nu \rho} \tilde{J}_i^{\mu \nu} \delta Y^{\rho}_i]
  \label{FJvara}
\end{equation}
where $ \tilde{J} _i=   p_i \delta_{\mathcal{D}_i}$ (with no sum over $i$).
This vanishes as a result of the veto condition (\ref{veto}). Then the
action indeed gives the desired field equations.
There is a similar dual theory for the dual potential $\tilde{A}$ with
$\tilde{F} = \ast J + d \tilde{A}$ in which $I_2 + I_3 $
is replaced with (\ref{duact}), and the analysis of this case is similar.

\section{$1$-form Symmetries of the Dirac Theory}\label{SymDir}

The smooth deformation of a Dirac string $\mathcal{D} $ to a Dirac string
$\mathcal{D} '$ while preserving the boundary $\partial \mathcal{D}$
(consisting of the particle worldline $\mathcal{C}$ together with a curve at
infinity $\mathcal{G}$, $\partial \mathcal{D}=\mathcal{C} \cup
\mathcal{G}$) gives a family of Dirac string world-sheets
$\mathcal{D} (\xi)$ parameterised by $\xi$ with $\mathcal{D} (0) =\mathcal{D}$
and $\mathcal{D} (1) =\mathcal{D}'$. This family of 2-dimensional world-sheets
sweeps out a 3-dimensional homotopy surface $\mathcal{E}$. It is important
that the deformation is consistent with the Dirac veto, so that, for every $\xi\in [0,1]$,  $\mathcal{D}
(\xi)$ intersects no particle worldline, in the case in which all particles are dyonic.
(For the case of magnetic monopoles and electric particles, if
the deformation is for the Dirac string of a magnetic monopole, then it should
not intersect the worldline of any particle carrying electric charge). If for
each $i$ the Dirac string $\mathcal{D}_i$ is deformed to $\mathcal{D}_i (\xi)$
sweeping out a 3-dimensional space $\mathcal{E}_i$, then the Dirac veto for $\mathcal{D}_i
(\xi)$ is
\begin{equation}
  \mathcal{D}_i (\xi) \cap \mathcal{C}_j = 0 \qquad \tmop{for} j \neq i
  \label{vetxi}
\end{equation}
This is required for all $\xi$ with $0 \leqslant \xi \leqslant 1$, so that the
homotpy surface $\mathcal{E}_i $ does not intersect any of the worldlines
$\mathcal{C}_j$ with $j \neq i$:
\begin{equation}
  \mathcal{E}_i \cap \mathcal{C}_j = 0 \qquad \tmop{for} j \neq i
  \label{defvet}
\end{equation}

For Dirac's case of magnetic monopoles and electric charges with veto (\ref{dirvetcase}), the
deformations $ \mathcal{D}_\alpha$ of the Dirac strings for the magnetic monopoles sweep out surfaces $\mathcal{E}_i $ and these are required not to intersect any of the worldlines 
 $\mathcal{C}_a$
of electrically charged particles, so that (\ref{defvet}) reduces to
\begin{equation}
  \mathcal{E}_\alpha \cap \mathcal{C}_a = 0  \end{equation}

The coordinates of $\mathcal{D} (\xi)$ are given by $x^{\mu} = Y ^{\mu} (\tau
, \sigma  ; \xi)$ for some function $Y ^{\mu} (\tau , \sigma  ; \xi)$ where
$\sigma^a = (\tau, \sigma)$ are the world-sheet coordinates. Then $\tau ,
\sigma , \xi$ are   coordinates for $\mathcal{E}$. From (\ref{FJvar})
\begin{equation}
  \int_{\mathcal{M}} F \wedge \delta \tilde{J} = p  \int_{\mathcal{D} } d^2
  \sigma 
  \,
   \epsilon^{a b} \, \partial_{[\rho } (\ast F)_{\mu 
  \nu]}  \frac{\partial Y^{\mu} }{\partial \sigma ^a} \frac{\partial Y^{\nu}
  }{\partial \sigma ^b} 
  \,
  \delta Y^{\rho} 
\end{equation}
under an infinitesimal change $Y \rightarrow Y + \delta Y$ with $\delta Y = 0$
on $\partial \mathcal{D}$. Let
\begin{equation}
  I (\xi) = \int_{\mathcal{M}} F \wedge \tilde{J} (\xi)
\end{equation}
Then
\begin{equation}
  \frac{d}{d \xi} I (\xi) = p  \int_{\mathcal{D} } d^2 \sigma 
  \,
   \epsilon^{a b}
   \,
  \partial_{[\rho } (\ast F)_{\mu  \nu]}  \frac{\partial
  Y^{\mu} }{\partial \sigma ^a} \frac{\partial Y^{\nu} }{\partial \sigma ^b}
  \frac{\partial Y^{\rho} }{\partial \xi}
\end{equation}
Integrating this over $\xi$ gives a change $\tilde{J} \rightarrow \tilde{J}'$
with
\begin{equation}
  \int_{\mathcal{M}} F \wedge (\tilde{J}' - \tilde{J}) = p 
  {\int_{\mathcal{M}}}_{} d (\ast F) \wedge \ast \delta_{\mathcal{E}} = p 
  \int_{ \mathcal{E}} d (\ast F) = p  \int_{ \mathcal{M}} F \wedge d^{\dag}
  \delta_{\mathcal{E}}
\end{equation}
so that
\begin{equation}
  \tilde{J}' - \tilde{J} = p d^{\dag} \delta_{\mathcal{E}}
\end{equation}
Then for finite shifts one has
\begin{equation}
  \rho = q \delta_{\mathcal{E}}, \quad \tilde{\rho} = p \delta_{\mathcal{E}}
\end{equation}
For an infinitesimal deformation $Y \rightarrow Y + \delta Y$, this gives
\begin{equation}
  \delta \tilde{J} = d^{\dag} \tilde{\rho}
  \label{varJ}
\end{equation}
where at $\xi = 0$
\begin{equation}
  \tilde{\rho}^{\mu \nu \rho} (x) = p  \int_{\mathcal{D} } d^2 \sigma
  \,
   \epsilon^{a b} \,
  \frac{\partial Y^{[\mu} }{\partial \sigma ^a} \frac{\partial Y^{\nu}
  }{\partial \sigma ^b} \frac{\partial Y^{\rho]} }{\partial \xi} \delta
  (x^{\lambda} - Y ^{\lambda} (\tau , \sigma  ; 0))
\end{equation}
This can be written as
\begin{equation}
  \tilde{\rho} = p \iota_V d \delta_{\mathcal{D}}
\end{equation}
where $V^{\rho}$=$\frac{\partial Y^{\rho} }{\partial \xi}$. Similarly, if $J =
q \delta_{\mathcal{D}}$, then
\begin{equation}
  \delta J = d^{\dag} \rho \delta \xi
\end{equation}
where
\begin{equation}
  \rho^{\mu \nu \rho} (x) = p \int_{\mathcal{D} } d^2 \sigma 
  \,
   \epsilon^{a b} \,
   \frac{\partial
  Y^{[\mu} }{\partial \sigma ^a} \frac{\partial Y^{\nu} }{\partial \sigma ^b}
  \frac{\partial Y^{\rho]} }{\partial \xi} \delta (x^{\lambda} - Y ^{\lambda}
  (\tau , \sigma  ; 0))
\end{equation}
Then an infinitesimal deformation of the Dirac string induces the
transformations (\ref{Jro}) with $\rho, \tilde{\rho}$ localised on the 
world-sheet $\mathcal{D}$ of the Dirac string.

\section{The Anomaly and the Veto}

Consider now the transformations (\ref{lam}),(\ref{lamt}) for general
parameters $\lambda, \tilde{\lambda}$, not necessarily restricted to be
associated with deformations of the Dirac strings. The field strength
(\ref{fja}) is invariant under (\ref{lam}) so that the change in the action
(\ref{act}) under (\ref{lam}) is
\begin{equation}
  \delta S = \int_{\mathcal{M}} \lambda \wedge \ast j = \sum_i q_i
  \int_{\mathcal{M}} \lambda \wedge \ast \delta_{\mathcal{C}_i} = \sum_i q_i
  \int_{\mathcal{C}_i} \lambda
\end{equation}
This variation will vanish if
\begin{equation}
   \lambda \wedge \ast j = 0 \label{vetls}
\end{equation}
and this condition  is equivalent to
\begin{equation}
  q_i \lambda \wedge \ast \delta_{\mathcal{C}_i} = 0 \label{vetl}
\end{equation}
for each $i$.
 The action (\ref{act}) does not involve $\tilde A$ or $J$ and so
is trivially invariant under the $\tilde{\lambda}$
transformations.

For a $\lambda$ arising from the change of $\tilde{J} = \sum_i p_i
\delta_{\mathcal{D}_i}$ that results from moving the Dirac strings one has
\begin{equation}
  \lambda = \sum_i p_i \ast \delta_{\mathcal{E}_i},  \label{varlam}
\end{equation}
so that (\ref{vetl}) becomes the condition that
\begin{equation}
  p_i q_j \delta_{\mathcal{E}_i} \wedge \delta_{\mathcal{C}_j} = 0
  \label{vetovero}
\end{equation}
for each $i, j$ (with no summation over $i$ or $ j$). If the particles are dyonic, then for $i \neq j$ this follows
from the Dirac veto while for $i = j$ this follows from the fact that
$\mathcal{C}_i $ is tangent to $\mathcal{E}_i$. This involves the
product of delta functions; similarly to the discussion in section 4, it is assumed that these are regularised by
replacing the delta functions with smooth functions in such a way that
$\delta_{\mathcal{E}_i} \wedge \delta_{\mathcal{C}_j} = 0$. For the case in
which all particles have either electric or magnetic charge but not both it
also follows from the fact that for each $i$ either $q_i $ or $p_i$ is zero. \

The alternative action (\ref{duact}) is invariant under (\ref{lam}) but under
(\ref{lamt}) it transforms as
\begin{equation}
  \delta S' = \int \tilde{\lambda} \wedge \ast \tilde{j} \label{spvar}
\end{equation}
This variation will vanish if
\begin{equation}
  p_i \tilde{\lambda} \wedge \ast \delta_{\mathcal{C}_i} = 0
\end{equation}
for each $i$. If $\tilde{\lambda}$ results from a deformation of the Dirac
strings, then it is of the form
\begin{equation}
\tilde{\lambda} = \sum_i q_i \ast \delta_{\mathcal{E}_i} \label{lamep}
\end{equation}
and the invariance condition
that (\ref{spvar}) vanishes is
\begin{equation}
  q_i p_j \delta_{\mathcal{E}_i} \wedge \delta_{\mathcal{C}_j} = 0
   \label{duveto}
\end{equation}
for all $i, j$. This is similar to (\ref{vetovero}), but with electric and magnetic charges interchanged.
As before, for $i \neq j$ this follows from the dual Dirac veto,
while for $i = j$ this follows from the fact that $\mathcal{C}_i $ is tangent
to $\mathcal{E}_i$.

\section{Dirac's Theory and Generalised Symmetries}

As discussed in the introduction, Dirac's theory can be related to the
generalised symmetry analysis of section 2 by defining 2-forms
\begin{equation}
  B = - \ast \tilde{J}, \quad \tilde{B} = - \ast J
  \label{sev1}
\end{equation}
together with field strengths
\begin{equation}
  H = d B = \ast \tilde{j}, \quad \tilde{H} = d \tilde{B} = \ast j
  \label{Hiss}
\end{equation}
Then $F = d A + \tilde{J}$ becomes
\begin{equation}
  F = d A - B \label{FABd}
\end{equation}
and the action
\begin{equation}
  S [A] = \int \frac{1}{2} F \wedge \ast F - A \wedge \ast j
\end{equation}
becomes
\begin{equation}
  S [A] = \int \frac{1}{2} F \wedge \ast F - A \wedge \tilde{H}
   \label{sev5}
\end{equation}
which, up to a surface term, can be rewritten as
\begin{equation}
  S [A] = \int \frac{1}{2} F \wedge \ast F + d A \wedge \tilde{B}
  \label{abact}
\end{equation}
Remarkably, this is precisely the action (\ref{actaa}) that arose from the
gauging of the 1-form symmetries in section 2. Adding a counterterm gives the
action (\ref{actbb}).

Writing
\[ \rho = \ast \tilde{\lambda}, \quad \tilde{\rho} = \ast \lambda \]
defines 1-form parameters $\lambda, \tilde{\lambda}$, and the transformations
(\ref{Jro}),(\ref{arho}) become
\begin{eqnarray}
  \delta A = \lambda, &  & \delta B = d \lambda  \label{lam2}\\
  \delta \tilde{A} = \tilde{\lambda}, &  & \delta \tilde{B} = d
  \tilde{\lambda}  \label{lamt2}
\end{eqnarray}
so that $B, \tilde{B}$ are gauge fields for the $\lambda, \tilde{\lambda}$
transformations and $F$ given by (\ref{FABd}) is invariant.

The action (\ref{abact}) is invariant under the $\tilde{\lambda}$
transformations but under the $\lambda$ transformations changes by
(\ref{anomaly}), which (up to a surface term) can be written as
\begin{equation}
  \delta S = - \int \lambda \wedge \tilde{H} \label{anomalyasd}
\end{equation}
so for a given
$\tilde{H}$, the system is invariant under transformations for which
(\ref{anomalyasd}) vanishes. For Dirac's theory,
\begin{equation}
\tilde{H} %\label{anomalyasd} 
=\ast  \sum_i q_i \delta_{\mathcal{C}_i}
\end{equation}
while $\lambda$ is given by (\ref{varlam}), so that the condition for
(\ref{anomalyasd}) to vanish is (\ref{vetovero}), which, as discussed in the
previous section, amounts to the Dirac veto condition.
Note that the variation (\ref{anomalyasd}) can be rewritten as
\begin{equation}
  \delta S = - \sum_i q_i  \int_{\mathcal{C}_i}  \lambda 
  \label{anomalyasdqq}
\end{equation}

Similarly, adding a counterterm to (\ref{abact}) gives the action
\begin{equation}
  S [A] = \int \frac{1}{2} F \wedge \ast F + F \wedge \tilde{B}
  \label{abactaa}
\end{equation}
which is invariant under the $\lambda$ transformations but under the
$\tilde{\lambda}$ transformations changes by (\ref{anomya}). For a given $\tilde{B}$,
the restriction to parameters $\tilde{\lambda}$ of the form (\ref{lamep}) for
which the variation vanishes is  (\ref{duveto}), which amounts to the dual
Dirac veto.

The Dirac veto can be written in the form (\ref{veto}), i.e.\ $\tilde{J} \wedge j = 0$.
Under the transformation (\ref{varJ}) this changes by
\begin{equation}
 \delta \tilde{J} \wedge j = d^{\dag} \tilde \rho  \wedge j =d^{\dag} (\tilde\rho  \wedge j)=0
 \label{vetod}
\end{equation}
which vanishes as a result of (\ref{vetls}).
Thus the Dirac veto condition is preserved under the generalised symmetries.

This gives an interesting reinterpretation of Dirac's theory. The theory can
be viewed as gauging the 1-form shift symmetries of Maxwell theory by coupling
to   2-form gauge fields $B, \tilde{B}$ that are chosen to be given
by the currents (\ref{Hiss}). For  given 2-form fields $B, \tilde{B}$,
the theory is invariant only under those gauge transformations for which
(\ref{anomaly}) or (\ref{anomya}) vanishes, giving a restriction on the gauge
parameters. In other words, the theory is invariant under those 1-form gauge
transformations for which the `anomaly' (\ref{anomaly}) or (\ref{anomya})
vanishes. For currents of the form (\ref{jiss}) and 1-form gauge parameters of the form
(\ref{varlam}),(\ref{lamep}) corresponding to changing the locations of the Dirac strings, this
restriction amounts to the Dirac veto. However, this reformulation allows a
generalisation to more general forms of the currents $j, \tilde{j}$ which then
lead to a corresponding modification of the restriction of the parameters
required for the anomaly to vanish, giving a generalisation of the veto.

\section{Field Equations, Quantum Symmetries and a Generalisation}

\subsection{Field Equations,  Symmetries and the Veto}

It was seen in section \ref{DiracTh} that the change in the action from varying the position of a Dirac string $Y^{\mu} (\tau, \sigma)\to Y^{\mu} (\tau, \sigma)+\delta Y^{\mu} (\tau, \sigma)$
is (\ref{FJvara}), which can be written as
\begin{equation}\delta S\propto  \int_{\mathcal{M}} \ast(
\epsilon_{\mu \nu \rho \sigma } j^\mu\tilde{J}  ^{ \nu\rho}   \delta Y^{\sigma} )
  \label{FJvarab}
\end{equation}
and this vanishes as a result of the veto condition (\ref{veto}).
It follows from the discussion in section \ref{SymDir}   that this can be rewritten as the change in the action from a generalised symmetry transformation $\delta \tilde{J} = d^{\dag} \tilde{\rho}
$   where
\begin{equation}
  \tilde{\rho}^{\mu \nu \rho} (x) = p  \int_{\mathcal{D} } d^2 \sigma
  \,
   \epsilon^{a b} \,
  \frac{\partial Y^{[\mu} }{\partial \sigma ^a} \frac{\partial Y^{\nu}
  }{\partial \sigma ^b} \delta   Y^{\rho]}  \delta
  (x^{\lambda} - Y ^{\lambda} (\tau , \sigma  ))
\end{equation}
Thus, as was to be expected, invariance of the action under the generalised symmetry transformations, which have been seen to correspond to deforming the positions of the Dirac strings, implies that the field equations are independent of the positions of the Dirac strings, and the Dirac veto is a sufficient condition for both.
However, the analysis of symmetries of the action is considerably simpler than that of the field equations in \ref{DiracTh}.

\subsection{The Quantum Theory}

As discussed in section \ref{GenSym}, 
 the condition for a classical symmetry is that it leaves the action invariant, 
while for a quantum theory it is    sufficient that the action be invariant modulo $2\pi \hbar$. %\footnote{For the Euclidean action, the condition is that it be invariant modulo $2\pi i\hbar$.}  
For the variation (\ref{anomaly}) (or, equivalently, (\ref{anomalyasd})) this gives the condition (\ref{anomalybb}):
\begin{equation}
  \int  \lambda \wedge \tilde{H} \in 2\pi \hbar \mathbb{Z} \label{anomalybbc}
\end{equation}
For Dirac's theory,
\begin{equation}
\tilde{H} %\label{anomalyasd} 
=\ast  \sum_i q_i \delta_{\mathcal{C}_i}
\,
,
\qquad
\lambda = \sum_i p_i \ast \delta_{\mathcal{E}_i}
\label{weweryg}
\end{equation}
giving the condition
\begin{equation}
 \sum_{i,j}  q_i  p_j\int 
 \delta_{\mathcal{C}_i}
   \wedge   \delta_{\mathcal{E}_j}\in 2\pi \hbar \mathbb{Z} \label{anomalybbcc}
\end{equation}
Here $ {\mathcal{E}_j}$ is a 3-surface swept out by
a family of Dirac string world-sheets
$\mathcal{D}_i (\xi)$
interpolating between Dirac strings ${\mathcal{D}_j}$
and  ${\mathcal{D}'_j}$. As discussed in   section \ref{SymDir}, if 
the
Dirac strings $\mathcal{D}_i
(\xi)$ all satisfy the Dirac veto (\ref{vetxi}), then
$ {\mathcal{E}_j}$ doesn't intersect any ${\mathcal{C}_i}$ with $i\ne j$
so that (\ref{defvet}) holds and (\ref{anomalybbcc}) is zero.

Consider now the more general case in which the initial and final Dirac strings ${\mathcal{D}_j}$
and  ${\mathcal{D}'_j}$ satisfy the Dirac veto, but the 3-surface $ {\mathcal{E}_j}$
intersects the curve ${\mathcal{C}_i}$ with $i\ne j$ in a finite number of points, given by an integer
\begin{equation}
 N_{ij} =\int _{\mathcal{M}} 
 \delta_{\mathcal{C}_i}
   \wedge   \delta_{\mathcal{E}_j}  \label{numby}
\end{equation}
Then the condition for a quantum symmetry (\ref{anomalybbcc}) follows from 
the Dirac quantisation condition (\ref{Diracquant}).
As the Dirac string is smoothly moved from ${\mathcal{D}_j}$
to  ${\mathcal{D}'_j}$ it crosses a finite number of  particle worldlines  ${\mathcal{C}_i}$ and with each crossing the action jumps by $2\pi \hbar $ times an integer but $\exp ( iS/\hbar)$ remains invariant. 

Dirac's action is then a multivalued function on the space of configurations  with a discontinuity at configurations not satisfying the Dirac veto, while $\exp ( iS/\hbar)$ is a well-defined continuous function on this space. In particular, the classical field equations are not well-defined at configurations not satisfying the veto, where the action is not continuous, reflecting the analysis  in section \ref{DiracTh}.
In the next section, the Dirac  theory will be embedded in a larger theory for which the action is a continuous function, and the anomaly is cancelled.

 Consider the Wilson line operator
 \begin{equation}
  W (\mathcal{C}_i) = \exp
 \left(  {\frac{i}{\hbar} q _i\int_{\mathcal{C}_i} A}\right) =
  \exp \left( {\frac{i}{\hbar} q _i\int_{\mathcal{M}} A\wedge \ast \delta_{\mathcal{C}_i} }\right)
\end{equation}
Under the shift $A\to A+\lambda$, this changes by
\begin{equation}
  W (\mathcal{C}_i) \to  \exp
 \left(  {\frac{i}{\hbar} q _i\int_{\mathcal{C}_i} \lambda}\right) W (\mathcal{C}_i)=
  \exp \left( {\frac{i}{\hbar} q _i\int_{\mathcal{M}} \lambda\wedge \ast \delta_{\mathcal{C}_i} }\right)W (\mathcal{C}_i)
\end{equation}
Using (\ref{weweryg}), the phase is
\begin{equation}
  \exp \left( {\frac{i}{\hbar} q _i\int_{\mathcal{M}} \lambda\wedge \ast \delta_{\mathcal{C}_i} }\right)  =\exp \left( {\frac{i}{\hbar} q _i\sum_j p_j\int_{\mathcal{M}} \delta_{\mathcal{E}_j} \wedge \ast \delta_{\mathcal{C}_i} }\right) %=\exp \left( {\frac{i}{\hbar} q _i\sum_j p_j N_{ij}  }\right)
  \end{equation}
  Then (\ref{numby}) implies this becomes
  \begin{equation}
  \exp \left( {\frac{i}{\hbar} q _i\sum_j p_j N_{ij}  }\right)=1  \end{equation}
which is equal to one as a result of the Dirac quantisation condition (\ref{Diracquant}).
Similarly, the 't Hooft lines
 \begin{equation}
 \tilde W (\mathcal{C}_i) = \exp
 \left(  {\frac{i}{\hbar} p _i\int_{\mathcal{C}_i} \tilde A}\right) \end{equation}
are invariant under the shift $\tilde A\to \tilde A+\tilde \lambda$.
Thus the generalised symmetries of the Dirac theory preserve the Wilson  and 't Hooft lines, so that they are observables for the theory.
  
 In conclusion, for both the classical and the quantum Dirac theory, the configurations are restricted to comply with the Dirac veto. The classical action is invariant under generalised symmetries corresponding to deforming the Dirac strings while preserving the Dirac veto, while the quantum theory has a larger invariance of deformations in which the interpolating strings $\mathcal{D}_i (\xi)$
 need not satisfy the veto, so long as  the initial and final  
 Dirac strings ${\mathcal{D}_j}$
and  ${\mathcal{D}'_j}$ obey the veto. 

\subsection{A Generalisation of the Dirac Model}

The Dirac model has sources consisting of the electric and magnetic 1-form currents $j,\tilde j$ localised on the particle worldlines,  while the 2-form current $\tilde J$ is  localised on the world-sheets of Dirac strings attached to the magnetically charged particles.
It is natural to seek a more general model   in which $j,\tilde j$ are taken to be  {\it any} conserved 1-form currents (not necessarily localised on particle worldlines) and $\tilde J$  is taken to be  {\it any} 2-form satisfying $d^\dagger \tilde J= \tilde j$.
 %Note that  $\tilde j$ needs to be conserved off-shell for such a  theory.
 Then this would be precisely the gauged Maxwell theory discussed in section 2
 with (\ref{sev1}),(\ref{Hiss}), field strength (\ref{FABd})  and action (\ref{sev5}) or (\ref{abact}), with generalised symmetries  (\ref{lamt2}). 
 Under a change of the choice of 
  $\tilde J$  satisfying $d^\dagger \tilde J= \tilde j$
  given by a generalised symmetry transformation $\tilde J\to \tilde J + d^\dagger \ast \lambda$, the action will be unchanged provided the 1-form  $\lambda$
    satisfies (\ref{anomalyaa}), while the condition for invariance of the quantum theory is (\ref{anomalybb}). 
This action gives the correct Maxwell equations (\ref{Ffida}).

A Dirac veto constraint could be imposed on this system in the form      (\ref{veto}) and this would be invariant under the generalised symmetries as a result of (\ref{vetod}).
   % If the currents  $j,\tilde j$ are treated as external sources, this provides a useful model. However, 
   If the currents  $j,\tilde j$ are constructed  from dynamical quantum particles or fields then $\tilde J$ also depends on the dynamical matter and can lead to some gauge dependence, non-locality  or ambiguity in the action for the quantum matter. Whether or not the veto is sufficient to eliminate these, as it does for the Dirac model, could depend on the details of the model.

\section{Anomaly Inflow and Dodging the Veto}

Intriguingly, the reformulation of the Dirac veto as a restriction to gauge
transformations for which the anomaly vanishes allows the cancellation of the
anomaly and hence the lifting of the veto in certain circumstances. For
example, if the four-dimensional spacetime $\mathcal{M}$ is the boundary of a
5-dimensional space $\mathcal{N}$ and if one introduces 2-forms $B',
\tilde{B}'$ on $\mathcal{N}$ that restrict to the 2-forms $B, \tilde{B}$
on $\mathcal{M}$, then the anomaly can be cancelled by adding the topological
action on $\mathcal{N}$ given by
\begin{equation}
  \int_{\mathcal{N}} B' \wedge d \tilde{B}'
  \label{topact}
\end{equation}
Extending the gauge fields $B, \tilde{B} $ to 5-dimensional fields $B',
\tilde{B}'$ on $\mathcal{N}$ corresponds to extending the sources to 5
dimensions. This can be done as follows. Each charged particle on
$\mathcal{M}$ is extended to a charged string in $\mathcal{N}$ that ends on
the particle on $\mathcal{M}$, so that each worldline $\mathcal{C}_i$ extends
to a world-sheet $\mathcal{C}_i'$ that ends on $\mathcal{M}$ in the worldline $\mathcal{C}_i$, i.e.
$\mathcal{C}_i' \cap \mathcal{M}= \mathcal{C}_i$. Attached to each particle on
$\mathcal{M}$ is a Dirac string on $\mathcal{M}$ with world-sheet $\mathcal{D}_i$. These are
extended to Dirac membranes $\mathcal{D}_i'$ in $\mathcal{N}$ ending on $\mathcal{M}$ in the
Dirac strings, \ i.e. $\mathcal{D}_i' \cap \mathcal{M}= \mathcal{D}_i$. The
 1-form currents (\ref{jiss}) then extend to 2-form currents
\begin{equation}
  j' = \sum_{i = 1}^N q_i \delta_{\mathcal{C}_i'}, \quad \tilde{j}' = \sum_{i
  = 1}^N p_i \delta_{\mathcal{C}_i'}
\end{equation}
and the 2-form current (\ref{Jtiis}) together with
\begin{equation}
 {J} = \sum_i q_i \delta_{\mathcal{D}_i}
  \label{Jtiisnot}
\end{equation}
extend to 3-form currents
\begin{equation}
  J' = \sum_{i = 1}^N q_i \delta_{\mathcal{D}_i'}, \quad \tilde{J}' = \sum_{i
  = 1}^N p_i \delta_{\mathcal{D}_i'}
\end{equation}
Then the 2-forms $B',\tilde{B}'$ on $\mathcal{N}$ are the 5-dimensional
duals of these
\begin{equation}
  B' = - \ast ' \tilde{J}', \quad \tilde{B}' = - \ast ' J'
\end{equation}
where   $\ast '$ denotes the Hodge dual on $\mathcal{N}$. Then the
5-dimensional interaction can be written as
\begin{equation}
  \int_{\mathcal{N}} j' \wedge \tilde{J}'
  \label{topact2}
\end{equation}
This means that the action given by this plus the Dirac action   on the boundary
$\mathcal{M}$ is formally invariant under the 1-form symmetries for which $B',
\tilde{B}'$ are gauge fields.   For the Dirac system, lifting the
particles in 4 dimensions to strings in 5 dimensions that couple to a gauge
field $A$ on the boundary is formally fully gauge invariant under the 1-form symmetries without needing to
impose a Dirac veto, provided the 5-dimensional action includes the coupling
between the electric and magnetic currents given by (\ref{topact2}). 

A similar but different way of lifting the theory to higher dimensions arises
when the 4-dimensional space  $\mathcal{M}$ is the world-volume of a 3-brane embedded in a higher dimensional spacetime.
 Consder the case of a  single D3-brane of type IIB string theory with
4-dimensional world-volume $\mathcal{M}$ embedded in a 10-dimensional space-time
$\mathcal{N}$. The world-volume theory involves a Maxwell field on  $\mathcal{M}$, with electrically charged particles at points
$r_i$ in $\mathcal{M}$ arising from fundamental strings in $\mathcal{N}$
ending at $r_i$ and magnetically charged particles at points $\tilde{r}_i$ in
$\mathcal{M}$ arising from  D-strings in $\mathcal{N}$ ending at
$\tilde{r}_i$. There is also a coupling of the 4-form gauge-field $C_4$ to the
D3-brane given by
\begin{equation}
  \int_{\mathcal{M}} C_4
\end{equation}
In this case, anomalous variation of the 4-dimensional theory under general
1-form symmetries can be cancelled by a suitable variation of $C_4$, again
giving a theory that is formally gauge-invariant without the need of demanding
the Dirac veto. This will be discussed further elsewhere.
 
\section{Generalisation to $p$-form gauge fields}

The generalisation to $p$-form gauge fields is straightforward, using the brane generalisation of Dirac's approach given in \cite{Deser:1997se,Deser:1997mz}. A $(p +
1)$-form field strength $F$ in a $d$ dimensional   spacetime satisfies the
equations
\begin{equation}
  dF = \ast \tilde{j}  \qquad d \ast F = \ast j \label{Ffidaa}
\end{equation}
where $j$ is a conserved $p$-form electric current and $\tilde{j}$ is a
conserved $\tilde{p}$-form magnetic current. If these currents arise from
electrically charged $p - 1$ branes with charges $q_i$ localised on
$p$-dimensional world-volumes $\mathcal{C}_i$ and magnetically charged
$\tilde{p} - 1$ branes with charges $p_{J}$ localised on
$\tilde{p}$-dimensional world-volumes $\tilde{\mathcal{C}}_{J}$ with
$\tilde{p} = d - p - 2$, then
\begin{equation}
  j (x) = \sum_i  q_i \delta_{\mathcal{C}_i} (x), \quad \tilde{j} (x) =
  \sum_{I}  p_{I} \delta_{\tilde{\mathcal{C}}_{I}} (x)
\end{equation}
The magnetic Dirac strings ending on monopole world-lines generalise to Dirac
$\tilde{p}$-branes with world-volumes $\tilde{\mathcal{D}}_I$ ending on
the magnetic $\tilde{p} - 1$ branes $\tilde{\mathcal{C}}_{I}$ while the
electric Dirac strings generalise to Dirac $p$-branes with world-volumes
$\mathcal{D}_i$ ending on the magnetic $p - 1$ branes $\mathcal{C}_i$. The
corresponding secondary currents
\begin{equation}
  J (x) = \sum_i  q_i \delta_{\mathcal{D}_i} (x), \quad \tilde{J} (x) =
  \sum_{I}  p_{I} \delta_{\tilde{\mathcal{D}}_{I}} (x)
\end{equation}
satisfy
\begin{equation}
  d^{\dag} J = j, \quad d^{\dag} \tilde{J} = \tilde{j}
\end{equation}
As before, the equation
\begin{equation}
  d F = \ast \tilde{j}
\end{equation}
can be solved by taking
\begin{equation}
  F = d A + \ast \tilde{J} \label{FAJ2}
\end{equation}
where now $A$ is a $p$-form potential. Dirac's action then generalises to $I =
I_1 + I_2 + I_3$ where
\begin{equation}
  I_1 = \sum_i T_i \int_{\mathcal{C}_i} L_i + \sum_{I}
  \tilde{T}_{I}\int_{\tilde{\mathcal{C}}}
  \tilde{L}_{I} \label{Act1}
\end{equation}
\begin{equation}
  I_2 = \int \frac{1}{2} F \wedge \ast F \label{Factid}
\end{equation}
\begin{equation}
  I_3 = \int A \wedge \ast j \label{Act3}
\end{equation}
where $L_i$ are the lagrangians for the electric $p - 1$ branes with tensions
$T_{i}$ and $\tilde{L}_{I}$ are the lagrangians for the magnetic
$\tilde{p} - 1$ branes with tensions $\tilde{T}_{I}$. These give the
correct field equations \cite{Dirac:1948um}, following the analysis in section \ref{DiracTh},
provided the veto condition
\begin{equation}
  \tilde{\mathcal{D}}_{J} \cap \mathcal{C}_i = 0
\end{equation}
holds, i.e.\ the magnetic Dirac $\tilde{p}$-branes do not intersect the
electric $p - 1$ branes. There is again a dual theory in terms of a dual
potential $\tilde{A}$ which is similar to the above but with the roles of
electric and magnetic sources interchanged.

The 1-form symmetries of the Maxwell theory generalise to $p$ and $\tilde{p}$
form symmetries (\ref{lam}),(\ref{lamt}) where now $\lambda$ is a $p$-form
parameter and $\tilde{\lambda}$ is a {$\tilde p$}-form parameter. As before, when
the parameters $\lambda, \tilde{\lambda}$ correspond to deformations of the
Dirac branes, then invariance under (\ref{lam}),(\ref{lamt}) follows provided
the field configurations are restricted to ones satisfying the veto condition.

\section{Born-Infeld Interactions and Self-Duality}

The $F^2$ kinetic term in (\ref{Facti}) or (\ref{Factid}) can be replaced by
a non-linear function $\mathcal{L} (F) $ of $F = d A + \ast J$, such as
the Born-Infeld lagrangian, so that (\ref{Facti}) or (\ref{Factid}) becomes
\begin{equation}
  I_2 = \int \ast \mathcal{L} (F) \label{BIf}
\end{equation}
Then, as shown in \cite{Dirac:1948um}, the action given by (\ref{BIf}) plus
(\ref{act1}),(\ref{act3}) or (\ref{Act1}),(\ref{Act3}) gives the desired field
equations.

If the rank $p + 1$ of $F$ is half the spacetime dimension, $p + 1 = d / 2$
(so that $p = \tilde{p}$) then one can consider self-dual field strengths $F =
\ast F$ in Lorentzian signature for even $p$ and in Euclidean signature for
odd $p$. The equations (\ref{Ffida}) are only consistent with $F = \ast F$ if
the electric and magnetic $p$-form currents are equal, $j = \tilde{j}$. The
corresponding $p - 1$ branes are then said to be self-dual. Then the equations
for a self-dual gauge field can be taken to be
\begin{equation}
  dF = \ast j, \quad F = \ast F \label{Ffidasd}
\end{equation}
as these then imply $d^{\dag} F = j$. The first equation then gives
\begin{equation}
  F = d A + \ast J
\end{equation}
as before. In addition,  there should be an action whose variation gives the self-duality condition. Some
such actions are discussed in \cite{Pasti:1996vs}-\cite{Evnin:2022kqn} and references therein.

\section{Extending the Wu-Yang approach}

\subsection{The Wu-Yang approach}

In the formalism of Wu and Yang \cite{Wu:1976qk}, the gauge field $A$ is a connection on a $U
(1)$ bundle over a submanifold $\bar{\mathcal{M}}$ of the spacetime
$\mathcal{M}$ on which $\tilde{j} = 0$. For example, $\bar{\mathcal{M}}$ could
be $\mathcal{M}$ with the world-lines of the magnetic monopoles and dyons
excised. Then on $\bar{\mathcal{M}}$ the absence of magnetic sources implies
$d F = 0$. For an open cover of $\bar{\mathcal{M}}$ with contractible patches
$\bar{U}_a$, on each patch $\bar{U}_a$ there will be a 1-form $A_a$ with $F =
d A_a$. As $\tilde{j} = 0$ on $\bar{\mathcal{M}}$, one can take $\tilde{J} =
0$ on $\bar{\mathcal{M}}$ and so there are no Dirac strings on $\bar{\mathcal{M}}$.

In each overlap
\begin{equation}
  A_a - A_b = d \sigma_{a b} \quad \tmop{in} \quad \bar{U}_a  \cap \bar{U}_b
  \label{atrans}
\end{equation}
for some 0-form $\sigma_{a b}$ on $\bar{U}_a  \cap \bar{U}_b$, and in triple
overlaps
\begin{equation}
  \sigma_{a b} + \sigma_{b c} + \sigma_{c a} = c_{a b c}\quad \tmop{in} \quad \bar{U}_a 
  \cap \bar{U}_b  \cap \bar U_c \label{lamtrip} 
  \end{equation}
for some constants $c_{a b c}$. Finally, for coupling to electrically charged particles
these constants are required to
satisfy 
\begin{equation}
  c_{a b c} \in 2 \pi \mathbb{Z} \label{cocy3}
\end{equation}
Then $A$ is a  connection on a $U(1)$ bundle over $\bar{\mathcal{M}}$ and
(\ref{cocy3}) is equivalent to the condition that $\frac{1}{2 \pi} F$ represents an
integral de Rham cohomology class.

This structure is important in defining $\int_{\mathcal{C}} A$. If the curve
$\mathcal{C}$ is contained in one patch $U_{\alpha}$, then $q
\int_{\mathcal{C}} A$ can be defined as $q \int_{\mathcal{C}} A_a$. If the
curve passes through several patches, however, there is a problem in defining
$q \int_{\mathcal{C}} A$ as $A$ is not a 1-form. For a closed curve that
bounds a 2-surface $\mathcal{D}$, 
the
definition of the coupling can be taken to be
$q \int_{\mathcal{D}} F$    as $F$ is a well-defined 2-form. As discussed in
section 3, this coupling depends on the choice of $\mathcal{D}$ but the
quantum theory is well-defined if the Dirac quantization condition holds.

For a curve that is not closed, such as the world-line of a particle, the
Wu-Yang coupling introduced in \cite{Wu:1976qk} can be used. For example, for a curve
$\mathcal{C}$ in $\bar{U}_a  \cup \bar{U}_b$ from $P \in \bar{U}_a $ to $Q \in
\bar{U}_b $ that passes through $\bar{U}_a  \cap \bar{U}_b$, the integral
along the curve is defined as
\begin{equation}
  \int_{P R} A_a + \sigma_{a b} (R) + \int_{R Q} A_b
\end{equation}
for any point $R \in \bar{U}_a  \cap \bar{U}_b$. The first integral is along
the part of the curve $\mathcal{C}$ from $P$ to $R$ and this lies entirely in
$\bar{U}_a $ and the final integral is along the part of the curve
$\mathcal{C}$ from $R$ to $Q$, and this lies entirely in $\bar{U}_b $. The
term $\sigma_{a b} (R)$ is needed to ensure that the result is independent of
the choice of mid-point $R \in \bar{U}_a  \cap \bar{U}_b$. This prescription
can be extended to any curve, but the extension to a curve that passes through
triple overlaps has a potential ambiguity which is avoided provided
(\ref{lamtrip}) holds \cite{Alvarez:1984es}. Finally the functional integral for an action
including this coupling is well-defined provided that the condition
(\ref{cocy3}) holds \cite{Alvarez:1984es}; this is the Dirac quantization condition.

The Wu-Yang action is then (\ref{act}) with $j$ given by (\ref{jiss}) so that
the coupling $\sum_i q_i \int_{\mathcal{C}_i} A$ defined according to the
Wu-Yang prescription  gives a well-defined quantum theory, provided that the
Dirac quantisation condition holds. Note that the charged particle
trajectories contributing to (\ref{act}) are all taken to lie in
$\bar{\mathcal{M}}$ and so they do not intersect any regions in which
$\tilde{j} \neq 0$.

\subsection{Combining the Wu-Yang and Dirac Approaches}

The Wu-Yang and Dirac approaches will be combined in the following to give a
treatment of the theory on the whole spacetime $\mathcal{M}$ that restricts to
the Wu-Yang formulation on $\bar{\mathcal{M}} \subset \mathcal{M}$.

A  manifold   requires multiple patches in general. 
Given a covering    $\{ \bar{U}_a \}$ of $\bar{\mathcal{M}} \subset
\mathcal{M}$, the covering of $\mathcal{M}$ can be chosen to be
$\{
U_{\alpha} \} = \{ \bar{U}_a, V_A \}$, consisting of the $\{ \bar{U}_a \}$ covering $\bar{\mathcal{M}}$ together with further patches $\{ V_A \}$ to extend the cover to
$\mathcal{M}$. 
All patches in the cover $\{
U_{\alpha} \} = \{ \bar{U}_a, V_A \}$ of $\mathcal{M}$ are taken to be contractible open
sets.  In this subsection it will be assumed that the magnetic current
$\tilde{j}$ is a smooth 1-form on $\mathcal{M}$ satisfying the conservation
law
\begin{equation}
  {d }^{\dag} \tilde{j} = 0
\end{equation}
Then the conservation of the magnetic current implies that in each
$U_{\alpha}$ there is a smooth 2-form $\tilde{J}_{\alpha}$ such that
\begin{equation}
  {d }^{\dag} \tilde{J}_{\alpha} = \tilde{j}
\end{equation}
Note that $\tilde{J}_{\alpha}$ is only non-zero in the patches $V_A$ and is
zero in the $\bar{U} $ that cover $\bar{\mathcal{M}}$.

Defining
\begin{equation}
  H = \ast \tilde{j}, \quad B_{\alpha} = \ast \tilde{J}_{\alpha}
\end{equation}
these can be written as
\begin{equation}
  H = d B_{\alpha}
\end{equation}
For a region $\mathcal{V}$ of a spacelike hypersurface bounded by a 2-surface
$\partial \mathcal{V}$, the magnetic charge contained in $\mathcal{V}$ is
\begin{equation}
  p = \int_{\mathcal{V}} \ast \tilde{j} = \int_{\mathcal{V}} H
\end{equation}
If $\mathcal{V}$ is contained in a single patch, $\mathcal{V} \subseteq
U_{\alpha}$ for some $U_{\alpha}$, then the magnetic charge can be written as
a surface integral
\begin{equation}
  p = \int_{\partial \mathcal{V}} B_{\alpha}
\end{equation}
More generally, the magnetic charge can be defined by the `Wilson surface'
$\int_{\partial \mathcal{V}} B $ which can be defined as in \cite{Alvarez:1984es},
provided that $B$ is a gerbe connection.

As $d F = \ast \tilde{j},$ it follows that in each $U_{\alpha}$
\begin{equation}
  d (F - B_{\alpha}) = 0
\end{equation}
so that there is a 1-form $A_{\alpha}$ on $U_{\alpha}$ with
\begin{equation}
  F = d A_{\alpha} + B_{\alpha}
\end{equation}
If $\mathcal{V}$ is contained in a single patch, $\mathcal{V} \subseteq
U_{\alpha}$ for some $U_{\alpha}$, then the magnetic charge can then be written in terms of $F$ in the standard way
\begin{equation}
  p = \int_{\partial \mathcal{V}} F
\end{equation}

In intersections of open sets the following relations apply
\begin{eqnarray}
  B_{\alpha} - B_{\beta} & =  d \zeta_{\alpha \beta} \qquad &\tmop{in} \quad
  U_{\alpha} \cap U_{\beta}
  \\
  \zeta_{\alpha \beta} + \zeta_{\beta \gamma} + \zeta_{\gamma \alpha} & =  d
  \Lambda_{\alpha \beta \gamma} \qquad &\tmop{in} \quad U_{\alpha} \cap
  U_{\beta} \cap U_{\gamma}\\
  \Lambda_{\alpha \beta \gamma} + \Lambda_{\beta \gamma \delta} +
  \Lambda_{\gamma \delta \beta} + \Lambda_{\delta \alpha \beta} & = 
  d_{\alpha \beta \gamma \delta} \qquad &\tmop{in} \quad U_{\alpha} \cap
  U_{\beta} \cap U_{\gamma} \cap U_{\delta}
 \end{eqnarray}
for a 1-form $\zeta_{\alpha \beta} $ in each intersection $U_{\alpha} \cap
U_{\beta}$, a 0-form $\Lambda_{\alpha \beta \gamma}$ in each triple
intersection $U_{\alpha} \cap U_{\beta} \cap U_{\gamma}$ and a constant
$d_{\alpha \beta \gamma \delta}$ in each quadruple intersection $U_{\alpha}
\cap U_{\beta} \cap U_{\gamma} \cap U_{\delta} $.
If the constants $d_{\alpha \beta \gamma \delta}$  satisfy the quantisation condition
\begin{equation}
  d_{\alpha \beta \gamma \delta}  \in  2 \pi \mathbb{Z}
\end{equation}
then $B$ is a gerbe connection and   $H$ represents an integral cohomology class, reflecting the quantisation of magnetic charge as $H=\ast \tilde j$.

Then in $U_{\alpha} \cap U_{\beta}$
\begin{equation}
  d (A_{\alpha} - A_{\beta}) = B_{\alpha} - B_{\beta} = d \zeta_{\alpha \beta}
\end{equation}
so
\begin{equation}
  A_{\alpha} - A_{\beta} = \zeta_{\alpha \beta} + d \sigma_{\alpha \beta}
  \label{atrans2}
\end{equation}
for some 0-form $\sigma_{\alpha \beta}$. In $U_{\alpha} \cap U_{\beta} \cap
U_{\gamma}$
\begin{equation}
  d (\sigma_{\alpha \beta} + \sigma_{\beta \gamma} + \sigma_{\gamma \alpha}) =
  - (\zeta_{\alpha \beta} + \zeta_{\beta \gamma} + \zeta_{\gamma \alpha}) = -
  d \Lambda_{\alpha \beta \gamma}
\end{equation}
so that
\begin{equation}
  \sigma_{\alpha \beta} + \sigma_{\beta \gamma} + \sigma_{\gamma \alpha} =
  c_{\alpha \beta \gamma} - \Lambda_{\alpha \beta \gamma} \label{strans2}
\end{equation}
for some constants $c_{\alpha \beta \gamma}$.

The action can be taken to be (\ref{act}) with $j$ given by (\ref{jiss}) with
the coupling $\sum_i q_i \int_{\mathcal{C}_i} A$ defined as above for curves
$\mathcal{C}_i$ lying in $\bar{\mathcal{M}} $ and hence satisfying the Dirac
veto. The theory has a generalised symmetry of the type discussed in previous
sections, with in each $U_{\alpha}$
\begin{equation}
  \delta A_{\alpha} = \lambda_{\alpha}, \quad \delta B_{\alpha} = d
  \lambda_{\alpha}
\end{equation}
for some 1-form $\lambda_{\alpha}$ on $U_{\alpha}$.

Recall that $\bar{\mathcal{M}} \subset
\mathcal{M}$ is a submanifold on which the magnetic currents vanish, $\tilde j \vert_{\bar{\mathcal{M}} }=0$.
A covering    $\{ \bar{U}_a \}$ of $\bar{\mathcal{M}} \subset
\mathcal{M}$ is extended to a covering  
$\{
U_{\alpha} \} = \{ \bar{U}_a, V_A \}$ of $\mathcal{M}$. Then on the patches $\{ \bar{U}_a \}$ there are no magnetic charges, so that $B_a = \ast \tilde{J}_{a}=0$ and hence $\zeta_{a b},
\Lambda_{a b c}, d_{a b c} $ can all be taken to be zero.
Then, on restricting to $\bar{\mathcal{M}} \subset \mathcal{M}$,
(\ref{atrans2}),(\ref{strans2}) reduce to (\ref{atrans}),(\ref{lamtrip})
and the Wu-Yang theory is recovered on  $\bar{\mathcal{M}}$.

\section{Discussion}

This paper has discussed the electromagnetic field interacting with
electrically charged particles, magnetic monopoles and dyons. This can arise as an
effective description of a gauge theory with gauge group spontaneously broken
to $U (1)$, with the 't Hooft-Polyakov monopoles approximated (at sufficient
distance from the monopole core) by Dirac monopoles. For heavy monopoles that
can be treated classically or semi-classically this can be regarded as a
theory of the electromagnetic field and electrically charged particles in a
magnetic monopole background, for which the Wu-Yang formulation can be used. However,
as has been discussed, such an approach is less useful when the magnetic
monopoles are light dynamical quantum particles. Such a scenario can arise,
for example, in $\mathcal{N}= 4$ super-Yang-Mills with coupling constant
$g_{\tmop{YM}} \sim 1$ and small Higgs vacuum expectation value  $v$ with W-bosons of mass $m_W \sim
v$ and magnetic monopoles of mass $m_M \sim v / g_{\tmop{YM}}^2$ both light
and of similar mass $m_W \sim m_M$. In the limit in which the Higgs vev goes to zero, $v \to 0$, both monopoles and W-bosons, together with a tower of dyons, become massless.

Dirac's theory is a quantum formulation of dynamical monopoles and electric
charges interacting with the electromagnetic field. It involves Dirac strings,
which are unobservable provided that the Dirac veto applies. Here it has been
shown that the Dirac string currents $J, \tilde{J}$ can be viewed as 2-form
gauge fields and that there are 1-form gauge symmetries associated with the
freedom in deforming the electric and magnetic Dirac strings. For Dirac's
theory, these are to be viewed as   gauge fields and the Dirac veto
is a restriction to backgrounds for which the mixed anomaly between the two
1-form symmetries vanishes.

There is, then, a close  relation between the gauged Maxwell theory discussed in section 2 and the Dirac theory,
but there is also an interesting difference.
In the discussion of section 2, the gauge fields $B,\tilde B$ were regarded as  gauge fields and were not integrated over. 
In the Dirac theory, the corresponding currents $J,\tilde J$ arise dynamically and are constructed from the worldlines of the quantum charged particles. 
However, these currents are not determined  completely
and it would be interesting to see if there was a formulation in which $B,\tilde B$  were integrated over, corresponding to integrating over the positions of the Dirac strings.
This would seem to require cancelling the anomaly or restricting to an integration over Dirac strings that abide by the veto.

In the theory considered here, the electric and magnetic 1-form currents $j,
\tilde{j}$ are conserved off-shell and so can be written as the divergences of
the 2-form currents $J, \tilde{J}$ without using field equations.
It would be interesting to seek a
generalisation of Dirac's theory that worked for cases in which the currents
$j, \tilde{j}$ are only conserved on-shell. 
One way of doing this could be to have $d^\dagger J=j$, $d^\dagger \tilde J=\tilde j$
arising as field equations for dynamical fields $J, \tilde{J}$.

It will be interesting to see if this could be a first step towards a generalisation to a theory in which the electric and magnetic sources are
fields, e.g.\  for the Maxwell field coupled to fermion fields with the electric
and magnetic currents given by fermion bilinears. Theories of this type have
that have been investigated previously by Schwinger \cite{Schwinger:1966nj} and Zwanziger \cite{Zwanziger:1970hk}
are non-local and have a number of
unusual features; see e.g.\  \cite{Blagojevic:1985sh} and references therein for further discussion.
The   treatment here could provide   an alternative approach.

\bigskip\bigskip
\noindent{\bf\Large Acknowledgements}:
\bigskip

\noindent 
I would like to thank Max Hutt, Rishi Mouland,  Marc Henneaux and Michele del Zotto for helpful discussions.
I would also like to thank Mohab Abou Zeid for a careful reading of the paper and useful comments.
 This research was supported by   the STFC Consolidated Grants   ST/T000791/1 and ST/X000575/1.


\begin{thebibliography}{99}

%\cite{Dirac:1948um}
\bibitem{Dirac:1948um}
P.~A.~M.~Dirac,
``The Theory of magnetic poles,''
Phys. Rev. \textbf{74} (1948), 817-830
doi:10.1103/PhysRev.74.817
%1174 citations counted in INSPIRE as of 24 Oct 2024

%\cite{Deser:1997se}
\bibitem{Deser:1997se}
S.~Deser, A.~Gomberoff, M.~Henneaux and C.~Teitelboim,
``P-brane dyons and electric magnetic duality,''
Nucl. Phys. B \textbf{520} (1998), 179-204
doi:10.1016/S0550-3213(98)00179-5
[arXiv:hep-th/9712189 [hep-th]].
%78 citations counted in INSPIRE as of 24 Oct 2024

%\cite{Deser:1997mz}
\bibitem{Deser:1997mz}
S.~Deser, A.~Gomberoff, M.~Henneaux and C.~Teitelboim,
``Duality, selfduality, sources and charge quantization in Abelian N form theories,''
Phys. Lett. B \textbf{400} (1997), 80-86
doi:10.1016/S0370-2693(97)00338-9
[arXiv:hep-th/9702184 [hep-th]].
%158 citations counted in INSPIRE as of 24 Oct 2024

%\cite{Deser:1997se},\cite{Deser:1997mz}

%\cite{Blagojevic:1985sh}
\bibitem{Blagojevic:1985sh}
M.~Blagojevic and P.~Senjanovic,
``The Quantum Field Theory of Electric and Magnetic Charge,''
Phys. Rept. \textbf{157} (1988), 233
doi:10.1016/0370-1573(88)90098-1
%106 citations counted in INSPIRE as of 16 Nov 2024


%\cite{Wu:1976qk}
\bibitem{Wu:1976qk}
T.~T.~Wu and C.~N.~Yang,
``Dirac's Monopole Without Strings: Classical Lagrangian Theory,''
Phys. Rev. D \textbf{14} (1976), 437-445
doi:10.1103/PhysRevD.14.437
%199 citations counted in INSPIRE as of 24 Oct 2024

%\cite{Alvarez:1984es}
\bibitem{Alvarez:1984es}
O.~Alvarez,
``Topological Quantization and Cohomology,''
Commun. Math. Phys. \textbf{100} (1985), 279
doi:10.1007/BF01212452
%171 citations counted in INSPIRE as of 01 Nov 2024



%\cite{Rham,Griffiths}
\bibitem{Rham}
G.~ de ~Rham, ``Differential manifolds. Forms, Currents, Harmonic Forms", Springer-Verlag 1984.
 
\bibitem{Griffiths}
P.~ Griffiths and J.~ Harris, ``Principles of Algebraic Geometry", John Wiley 1978.



%\cite{Lechner:2000pn}
\bibitem{Lechner:2000pn}
K.~Lechner and P.~Marchetti,
``Dirac branes, characteristic currents and anomaly cancellations in five-branes,''
Nucl. Phys. B Proc. Suppl. \textbf{102} (2001), 94-99
doi:10.1016/S0920-5632(01)01542-0
[arXiv:hep-th/0103161 [hep-th]].
%10 citations counted in INSPIRE as of 02 May 2025

%\cite{Cariglia:2004ez}
\bibitem{Cariglia:2004ez}
M.~Cariglia and K.~Lechner,
%`Intersecting D-branes, Chern-kernels and the inflow mechanism,''
Nucl. Phys. B \textbf{700} (2004), 157-182
doi:10.1016/j.nuclphysb.2004.08.023
[arXiv:hep-th/0406083 [hep-th]].
%2 citations counted in INSPIRE as of 02 May 2025

%\cite{Frey:2019fqz}
\bibitem{Frey:2019fqz}
A.~R.~Frey,
``Dirac branes for Dirichlet branes: Supergravity actions,''
Phys. Rev. D \textbf{102} (2020) no.4, 046017
doi:10.1103/PhysRevD.102.046017
[arXiv:1907.12755 [hep-th]].
%3 citations counted in INSPIRE as of 02 May 2025


%\cite{Lechner:2000pn,Cariglia:2004ez,Frey:2019fqz}




%\cite{Dirac:1931kp}
\bibitem{Dirac:1931kp}
P.~A.~M.~Dirac,
``Quantised singularities in the electromagnetic field,,''
Proc. Roy. Soc. Lond. A \textbf{133} (1931) no.821, 60-72
doi:10.1098/rspa.1931.0130
%2628 citations counted in INSPIRE as of 24 Oct 2024

%\cite{Bhardwaj:2023kri,Brennan:2023mmt,Schafer-Nameki:2023jdn}

%\cite{Gaiotto:2014kfa}
\bibitem{Gaiotto:2014kfa}
D.~Gaiotto, A.~Kapustin, N.~Seiberg and B.~Willett,
``Generalized Global Symmetries,''
JHEP \textbf{02} (2015), 172
doi:10.1007/JHEP02(2015)172
[arXiv:1412.5148 [hep-th]].
%1241 citations counted in INSPIRE as of 01 Nov 2024

%\cite{Bhardwaj:2023kri}
\bibitem{Bhardwaj:2023kri}
L.~Bhardwaj, L.~E.~Bottini, L.~Fraser-Taliente, L.~Gladden, D.~S.~W.~Gould, A.~Platschorre and H.~Tillim,
``Lectures on generalized symmetries,''
Phys. Rept. \textbf{1051} (2024), 1-87
doi:10.1016/j.physrep.2023.11.002
[arXiv:2307.07547 [hep-th]].
%128 citations counted in INSPIRE as of 01 Nov 2024

%\cite{Brennan:2023mmt}
\bibitem{Brennan:2023mmt}
T.~D.~Brennan and S.~Hong,
``Introduction to Generalized Global Symmetries in QFT and Particle Physics,''
[arXiv:2306.00912 [hep-ph]].
%106 citations counted in INSPIRE as of 01 Nov 2024

%\cite{Schafer-Nameki:2023jdn}
\bibitem{Schafer-Nameki:2023jdn}
S.~Schafer-Nameki,
``ICTP lectures on (non-)invertible generalized symmetries,''
Phys. Rept. \textbf{1063} (2024), 1-55
doi:10.1016/j.physrep.2024.01.007
[arXiv:2305.18296 [hep-th]].
%169 citations counted in INSPIRE as of 01 Nov 2024


%\cite{Abbott:1981ff}
\bibitem{Abbott:1981ff}
L.~F.~Abbott and S.~Deser,
``Stability of Gravity with a Cosmological Constant,''
Nucl. Phys. B \textbf{195} (1982), 76-96
doi:10.1016/0550-3213(82)90049-9
%1002 citations counted in INSPIRE as of 01 Nov 2024

%\cite{Hull:2023iny}
\bibitem{Hull:2023iny}
C.~M.~Hull,
``Magnetic charges for the graviton,''
JHEP \textbf{05} (2024), 257
doi:10.1007/JHEP05(2024)257
[arXiv:2310.18441 [hep-th]].
%6 citations counted in INSPIRE as of 01 Nov 2024

%\cite{Hull:2024bcl}
\bibitem{Hull:2024bcl}
C.~Hull, M.~L.~Hutt and U.~Lindstr\"om,
``Generalised symmetries in linear gravity,''
[arXiv:2409.00178 [hep-th]].
%2 citations counted in INSPIRE as of 05 Nov 2024

%\cite{Pasti:1996vs}
\bibitem{Pasti:1996vs}
P.~Pasti, D.~P.~Sorokin and M.~Tonin,
``On Lorentz invariant actions for chiral p forms,''
Phys. Rev. D \textbf{55} (1997), 6292-6298
doi:10.1103/PhysRevD.55.6292
[arXiv:hep-th/9611100 [hep-th]].
%332 citations counted in INSPIRE as of 05 Nov 2024

%\cite{Sen:2015nph}
\bibitem{Sen:2015nph}
A.~Sen,
``Covariant Action for Type IIB Supergravity,''
JHEP \textbf{07} (2016), 017
doi:10.1007/JHEP07(2016)017
[arXiv:1511.08220 [hep-th]].
%75 citations counted in INSPIRE as of 05 Nov 2024

%\cite{Sen:2019qit}
\bibitem{Sen:2019qit}
A.~Sen,
``Self-dual forms: Action, Hamiltonian and Compactification,''
J. Phys. A \textbf{53} (2020) no.8, 084002
doi:10.1088/1751-8121/ab5423
[arXiv:1903.12196 [hep-th]].
%71 citations counted in INSPIRE as of 05 Nov 2024


%\cite{Hull:2023dgp}
\bibitem{Hull:2023dgp}
C.~M.~Hull,
``Covariant action for self-dual p-form gauge fields in general spacetimes,''
JHEP \textbf{04} (2024), 011
doi:10.1007/JHEP04(2024)011
[arXiv:2307.04748 [hep-th]].
%14 citations counted in INSPIRE as of 05 Nov 2024

%\cite{Avetisyan:2022zza}
\bibitem{Avetisyan:2022zza}
Z.~Avetisyan, O.~Evnin and K.~Mkrtchyan,
``Nonlinear (chiral) p-form electrodynamics,''
JHEP \textbf{08} (2022), 112
doi:10.1007/JHEP08(2022)112
[arXiv:2205.02522 [hep-th]].
%25 citations counted in INSPIRE as of 05 Nov 2024

%\cite{Evnin:2022kqn}
\bibitem{Evnin:2022kqn}
O.~Evnin and K.~Mkrtchyan,
``Three approaches to chiral form interactions,''
Differ. Geom. Appl. \textbf{89} (2023), 102016
doi:10.1016/j.difgeo.2023.102016
[arXiv:2207.01767 [hep-th]].
%25 citations counted in INSPIRE as of 05 Nov 2024


%\cite{Pasti:1996vs}-\cite{Evnin:2022kqn}



%\cite{Schwinger:1966nj}
\bibitem{Schwinger:1966nj}
J.~S.~Schwinger,
``Magnetic charge and quantum field theory,''
Phys. Rev. \textbf{144} (1966), 1087-1093
doi:10.1103/PhysRev.144.1087
%495 citations counted in INSPIRE as of 16 Nov 2024

%\cite{Zwanziger:1970hk}
\bibitem{Zwanziger:1970hk}
D.~Zwanziger,
``Local Lagrangian quantum field theory of electric and magnetic charges,''
Phys. Rev. D \textbf{3} (1971), 880
doi:10.1103/PhysRevD.3.880
%357 citations counted in INSPIRE as of 16 Nov 2024



\end{thebibliography}
\end{document}